\documentclass[
aip,
 amsmath,amssymb,
preprint,
]{revtex4-1}

\usepackage{graphicx}
\usepackage{dcolumn}
\usepackage{bm}

\usepackage[utf8]{inputenc}
\usepackage[T1]{fontenc}
\usepackage{mathptmx}
\usepackage{etoolbox}
\usepackage{multirow}

\makeatletter
\def\@email#1#2{%
 \endgroup
 \patchcmd{\titleblock@produce}
  {\frontmatter@RRAPformat}
  {\frontmatter@RRAPformat{\produce@RRAP{*#1\href{mailto:#2}{#2}}}\frontmatter@RRAPformat}
  {}{}
}%
\makeatother

\usepackage{xcolor}

\usepackage{soul} 

\newcommand{\commentout}[1]{}

\newcommand{\OhNum}{{\mathrm{Oh}}}
\newcommand{\kB}{{k_\mathrm{B}}}

\newcommand{\SpeciesFlux}{\mathcal{F}}
\newcommand{\ReversibleStress}{\mathcal{R}}
\newcommand{\WhiteNoiseMass}{\mathcal{Z}}
\newcommand{\WhiteNoiseMomentum}{\mathcal{W}}
\newcommand{\ViscousTensor}{\boldsymbol{\tau}}

\newcommand{\StaticCA}{\theta_\mathrm{s}} 
\newcommand{\DiffE}{D_\mathrm{E}}   
\newcommand{\DiffSE}{D_\mathrm{SE}} 
\newcommand{\DiffB}{D_\mathrm{B}}   

\begin{document}

\preprint{AIP/123-QED}

\title[Droplet Dynamics by Fluctuating Hydrodynamics]
{A Study of Spherical and Sessile Droplet Dynamics by Fluctuating Hydrodynamics}
\author{John B. Bell*}\email{jbbell@lbl.gov} 
\author{Andrew Nonaka}%
\affiliation{ 
Lawrence Berkeley National Laboratory
}%

\author{Alejandro L. Garcia}
\affiliation{%
Dept. of Physics \& Astronomy, San Jose State Univ.}%

\date{\today}

\begin{abstract}
We simulate the mesoscopic dynamics of droplets formed by phase separated fluids at nanometer scales where thermal fluctuations are significant. Both spherical droplets fully immersed in a second fluid and sessile droplets which are also in contact with a solid surface are studied.
%
%
Our model combines a Cahn-Hillard formulation with incompressible fluctuating hydrodynamics; for sessile droplets the fluid-solid contact angle is specified as a boundary condition.
Deterministic simulations with an applied body force are used to measure the droplets' mobility from which a diffusion coefficient is obtained using the Einstein relation.
Stochastic simulations are independently used to obtain a diffusion coefficient from a linear fit of the variance of a droplet's position with time.
In some scenarios these two measurements give the same value but not in the case of a spherical droplet initialized near a slip wall or in the case of sessile droplets with large contact angles ($\geq 90^\circ$) on both slip and no-slip surfaces.


\end{abstract}

\maketitle

\section{Introduction}

The study of microscopic fluid droplets has received considerable attention\cite{bonn2009wetting,lohse2015surface,Qian2019nanodroplets} 
due to its importance in applications such as additive manufacturing\cite{sun2023droplet}, heat transfer\cite{DropJumping2020,DropBoiling_2022},
biosensors\cite{zhu2013analytical,qiao2014oil} and,
hydrogen storage\cite{HydrogenBubble2015}.
Additionally, the interaction between nano-droplets and solid surfaces has been a subject of interest, with phenomena like wetting, spreading, and adhesion exhibiting unique features at near-molecular scales.
Most of these studies have considered either the dynamics of spherical droplets floating in a medium or static sessile droplets on surfaces.
There have been a number of nanoscale fluid simulations of surface wetting\cite{WettingMD_1999,WettingMD_2009} yet the Brownian motion of surface droplets has been investigated in only a few molecular dynamics\cite{DropDiffusionMD_2016,DropDiffusionMD_2018,DropMergeMD_2022} and
Multiparticle Dissipative Particle Dynamic\cite{chang2016wetting,chang2016resisting} studies.

In this paper we use a multiphase fluctuating hydrodynamic (FHD) model\cite{MultiphaseFHD2014,RTIL,Gallo2022,RayPlatFHD} to simulate fluid droplets in a fluid medium both on and near solid surfaces. 
The theoretical foundation of our model is the stochastic extension of the Navier-Stokes equations introduced by Landau and Lifshitz\cite{Landau_59,Zarate_07}.
In this regard it is similar to stochastic lubrication theory, which has been successful in describing capillary wave fluctuations and instabilities\cite{ThinFilmMecke2001,ThinFilmGrun2006,CapWaveThinFilm_2021,ThinFilmFluct_2021,CapWaveFluct2023},
however, our FHD formulation does not assume a thin film covering the entire surface.

Our fluctuating multiphase formulation is based on the Cahn-Hillard model\cite{CahnHilliard:1958,anderson1998diffuse} that describes a phase separated, binary liquid mixture.
The deterministic form of this model is commonly applied to the
study of droplet dynamics (e.g., surface impact\cite{khatavkar2007diffuse})
and deterministic Cahn-Hillard Navier-Stokes calculations of droplet coalescence are found to be in excellent agreement with molecular dynamics simulations over the entire coalescence process, including the decay of the inertia-induced oscillation of the merged droplet~\cite{Droplets_MD_CH_2022}.
Our model uses a fluid-solid boundary condition derived from a surface free energy that specifies contact angle.\cite{ContactAngleBCs2000,ContactAngleBCs2012}
As originally shown by Seppecher~\cite{Seppecher1996moving,Sibley2013contactline}, in the diffuse interface model sessile droplets are not pinned because there is no dynamical singularity near the contact line, in contrast to a sharp interface model that has a singularity where the interface intersects the boundary.\cite{Huh1971hydrodynamic,bonn2009wetting,Snoeijer2013moving}

The next section outlines the model, followed by a description of the algorithm and its validation. 
Numerical results for the mobility and diffusion coefficient of spherical and sessile droplets are then presented for a variety of scenarios. 
We report the conditions for which the diffusion coefficient obtained from Einstein relation using the measured mobility matches the value from the Brownian motion of the droplets as measured in the simulations as well as conditions in which they differ.
We conclude with a summary of the current work and potential future studies.

\section{Fluctuating Hydrodynamics}

\subsection{Theory}

Our multispecies fluctuating dynamics formulation is similar to that described in Barker \textit{et al.}\cite{BrynRP2023}.
Consider a binary mixture of similar species, A and B, with a specific free energy density,\cite{CahnHilliard:1958}
\begin{align}
\frac{\mathcal{G}}{\rho \kB T} =
c \ln c + (1-c) \ln (1-c) 
+ \chi c (1-c) 
+  \kappa |\nabla c |^2 
\label{eq:CHfree} 
\end{align}
where $\rho$ is the mass density, $c$ is the mass fraction of species B, $T$ is temperature,
and $\kB$ is the Boltzmann constant.
For interaction coefficient $\chi \leq 2$ the mixture is homogeneous at equilibrium; for $\chi > 2$ it phase separates into concentrations $c_{e,1}$ and $c_{e,2} = 1 - c_{e,1}$ given by solutions of
\begin{align}
\ln \left( \frac{c_e}{1-c_e}\right) = \chi (2 c_e - 1) 
\label{eq:ce_eqn}
\end{align}
The surface energy coefficient is $\kappa$ and the surface tension is
\begin{align}
    \gamma = n \kB T \sqrt{2 \chi \kappa} ~\sigma_r(\chi)
    \label{eq:SurfaceTension}
\end{align}
where $n = \rho/m$ is the number density and $\sigma_r \simeq O(1)$.
The characteristic length scale for the interface is $\ell_\mathrm{c} = \sqrt{2\kappa / \chi}$ and for capillary wave fluctuations is $\ell^\star = \sqrt{ \kB T / \gamma}$.

The incompressible flow equations for constant $\rho$ are
\begin{align}
( \rho c)_t + \nabla \cdot(\rho u c) =& \nabla \cdot {\SpeciesFlux}   \nonumber \\
( \rho u)_t + \nabla \cdot(\rho u u) + \nabla \pi =& \nabla \cdot {\ViscousTensor}  + \nabla \cdot \ReversibleStress \nonumber \\
\nabla \cdot u =& 0
\label{eq:low_mach_eqs}
\end{align}
where $u$ is the fluid velocity and $\pi$ is a perturbational pressure.
Here, $\SpeciesFlux$, $\ViscousTensor$, and $\ReversibleStress$ are the species flux, viscous stress tensor, and the reversible stress due to the interfacial tension, respectively.

In fluctuating hydrodynamics the dissipative fluxes are written as the sum of deterministic and stochastic terms. 
The species flux is $\SpeciesFlux = \overline{\SpeciesFlux} + \widetilde{\SpeciesFlux}$ where the deterministic flux is
\begin{align}
\label{eq:DetSpeciesFlux}
\overline{\SpeciesFlux} = \rho D \left ( \nabla c - 2 \chi  c (1-c) \nabla c +  2c(1-c) \kappa \nabla \nabla^2 c \right )
\end{align}
and $D$ is the species diffusion coefficient. 
The stochastic flux is
$\widetilde{\SpeciesFlux} = \sqrt{2 \rho m D c (1-c)}  ~\WhiteNoiseMass$
where $\WhiteNoiseMass(\mathbf{r},t)$ is an uncorrelated Gaussian random vector vector field with covariance $\delta_{\alpha,\beta} \delta(t-t') \delta (r-r-') $.
The viscous incompressible stress tensor is
$\ViscousTensor =  \overline{\ViscousTensor} +  \widetilde{\ViscousTensor}$
where the deterministic component is
$\overline{\ViscousTensor} = \eta [\nabla u + (\nabla u)^T]$.
The stochastic contribution to the viscous stress tensor is 
$\widetilde\ViscousTensor = \sqrt{\eta \kB T}({\WhiteNoiseMomentum} + {\WhiteNoiseMomentum}^T),
$ where 
${\WhiteNoiseMomentum}(\mathbf{r},t)$ is a standard Gaussian random tensor field
with uncorrelated components.
Finally, the interfacial reversible stress is
\begin{align}
\ReversibleStress = n \kB T \kappa \left [\frac{1}{2} |\nabla c|^2 \mathbb{I} - \nabla c \otimes \nabla c \right].
\label{eq:reversible_stress}
\end{align}
Note that since $\ReversibleStress$ is a non-dissipative flux there is no corresponding stochastic flux.

The solid-fluid boundary condition is derived by introducing a surface free energy as done by 
Jacqmin\cite{ContactAngleBCs2000} and Dong\cite{ContactAngleBCs2012} to obtain
\begin{align}
    \hat{n}\cdot \nabla c = \frac{3 m \gamma \cos \StaticCA}{\rho \kB T \kappa} \frac{(c-c_{e,1})(c_{e,2}-c)}{(c_{e,2}-c_{e,1})^3}
    \label{eq:ContactAngleBC}
\end{align}
where $\StaticCA$ is the static contact angle. 
Our construction parallels the derivation of Dong\cite{ContactAngleBCs2012} modified to reflect that, in the absence of noise, the concentration transitions from $c = c_{e,1}$ in the far field of the droplet to $c = c_{e,2}$ inside the droplet (see Eq.~\ref{eq:ce_eqn}).
We do not consider the contribution of disjoining pressure, which affects the contact line at a scale that is smaller than that resolved in our simulations.

\subsection{Numerical Formulation}

The system of equations [\ref{eq:low_mach_eqs}] is discretized using a structured-grid finite-volume approach with cell-averaged concentrations and face-averaged (staggered) velocities with standard spatial discretizations. The algorithm uses an explicit discretization of concentration coupled to a semi-implicit discretization of velocity using a predictor-corrector scheme for second-order temporal accuracy.  
The momentum equation is discretized using a Stokes-type splitting.  Specifically, the advective terms and the reversible stress are computed explicitly using data at time $t^n$ in the predictor and $t^{n+1}$ in the corrector.  These terms then form part of the right hand side of a Stokes system that treats the viscous tensor and the constraint implicitly.
The discretized Stokes system is solved by a generalized minimal residual (GMRES) method with a multigrid preconditioner, see Cai \textit{et al.}\cite{cai:2014}.
The explicit treatment of the concentration equation introduces a stability limitation on the time step of
\begin{align}
D \left( \frac{12}{\Delta x^2} + \frac{72 \kappa}{\Delta x^4} \right) \Delta t \leq 1
\label{eq:StableDt}
\end{align}
where $\Delta x$ is the mesh spacing.
The numerical scheme is based on methods introduced in earlier work~\cite{donev2014low, Donev_10, RTIL}; details are discussed in Barker \textit{et al.}\cite{BrynRP2023} and its Supporting Information.
The computational model has been validated on simulations of capillary waves generated by thermal fluctuations \cite{bell2024commentbrownianmotiondroplets}.  In particular, the
simulations showed good agreement with stochastic lubrication theory and molecular dynamics simulations.

\section{Droplet Simulations}

Unless otherwise specified, the physical parameters used in all simulations are as follows: 
mass density, $\rho = 1.0~\mathrm{g/cm}^3$, 
molecular mass, $m = 6.0\times 10^{-23}$~g, 
temperature $T = 100$K. 
The Flory interaction parameter $\chi = 3.5$, 
which corresponds to a reduced temperature of about 0.57;
the equilibrium concentrations are $c_{e,1} = 0.037874$ and $c_{e,2} = 1 - c_{e,1} = 0.962126$.
The surface energy coefficient is $\kappa = 3.0\times 10^{-14}~\mathrm{cm}^{2}$
giving a surface tension, $\gamma = 24.3386$~dyne/cm.
For these values $\ell_\mathrm{c} = 1.31$~nm, $\ell_* = 0.238$~nm, and
the interface thickness is roughly $2.6 \ell_\mathrm{c} \approx 3.45$~nm.
The viscosity and the species diffusion coefficient are
$\eta = 0.01$~poise and $D = 2 \times 10^{-5}~\mathrm{cm^2/s}$, independent of concentration.  
For these parameters, the Ohnesorge number, which compares viscous forces to inertial and surface tension forces, is $\OhNum = \eta/\sqrt{\rho R \gamma} \approx 2$ for a sphere of radius $R = 10$~nm.
In a typical simulation the droplet represents approximately $7 \times 10^4$ molecules of species 2 in a domain with roughly $3 \times 10^7$ molecules of species 1.

The simulation mesh was a grid with $N_x \times N_y \times N_z$ grid points with spacings $\Delta x = \Delta y = \Delta z = 1.0$~nm.
In general, simulations used $N_x = N_y = 128$; for 2D cases $N_z = 1$ and for 3D cases $N_z = 128$.
The time step was $\Delta t = 0.6$~ps, which corresponds to approximately $1/3$ of the maximum stable time step (see Eq.~\ref{eq:StableDt}).

\begin{figure}[h!]
  \centering
      \includegraphics[width=0.85\textwidth]{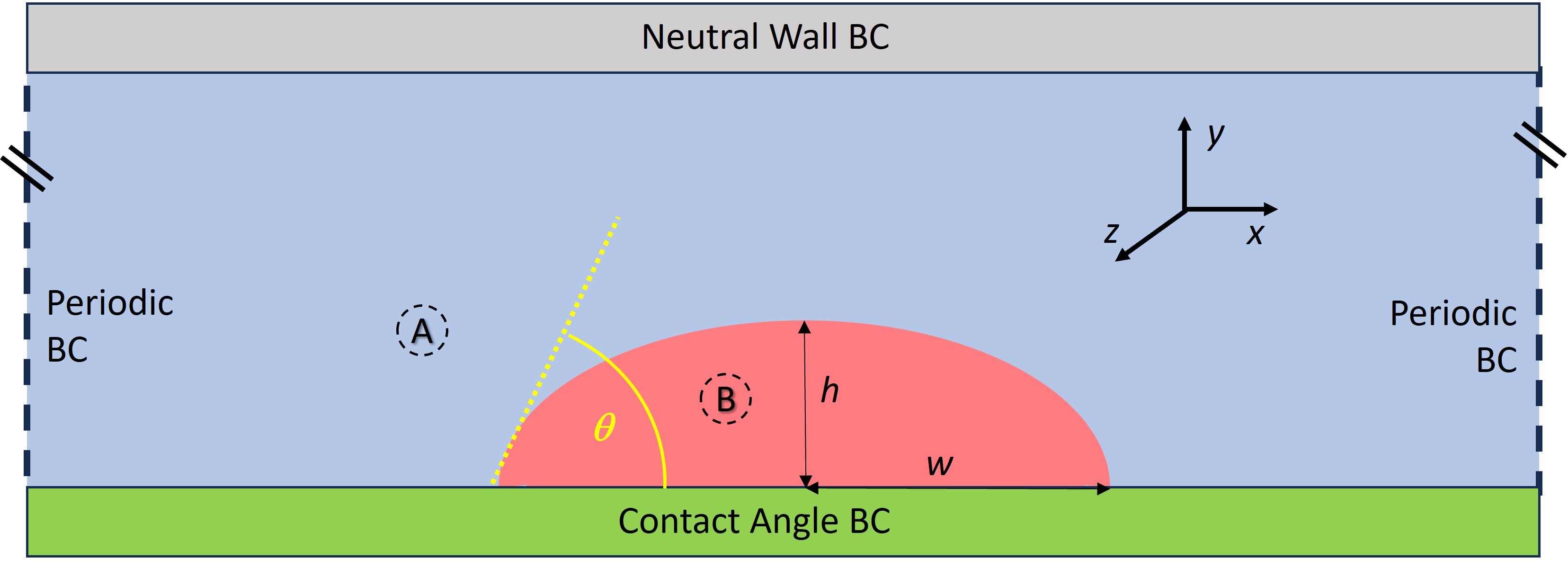}
    \caption{Simulation geometry for sessile droplets on a flat, solid surface (plane at $y=0$).}
    \label{fig:SimGeometry}
\end{figure}

Figure~\ref{fig:SimGeometry} illustrates the geometry used for most of the simulations, the exception being when all boundaries were periodic. 
For all cases the boundary conditions in $x$ and $z$ directions are periodic.
For a spherical droplet in bulk fluid the boundary condition in the $y$ direction is also periodic.
In all other cases the $y=0$ plane is a contact angle boundary and the $y=L_y$ plane is a neutral wall boundary (see Fig.~\ref{fig:SimGeometry}).
The contact angle boundary condition imposes the concentration boundary condition (see Eq.~\ref{eq:ContactAngleBC}).
The neutral wall boundary condition is similar but always with $\StaticCA = 90^\circ$ boundary conditions ($\hat{n}\cdot \nabla c = 0$). 
Velocity on both walls are treated with either slip or no-slip boundary conditions.

\begin{figure}
  \centering
    \includegraphics[width=0.40\textwidth]{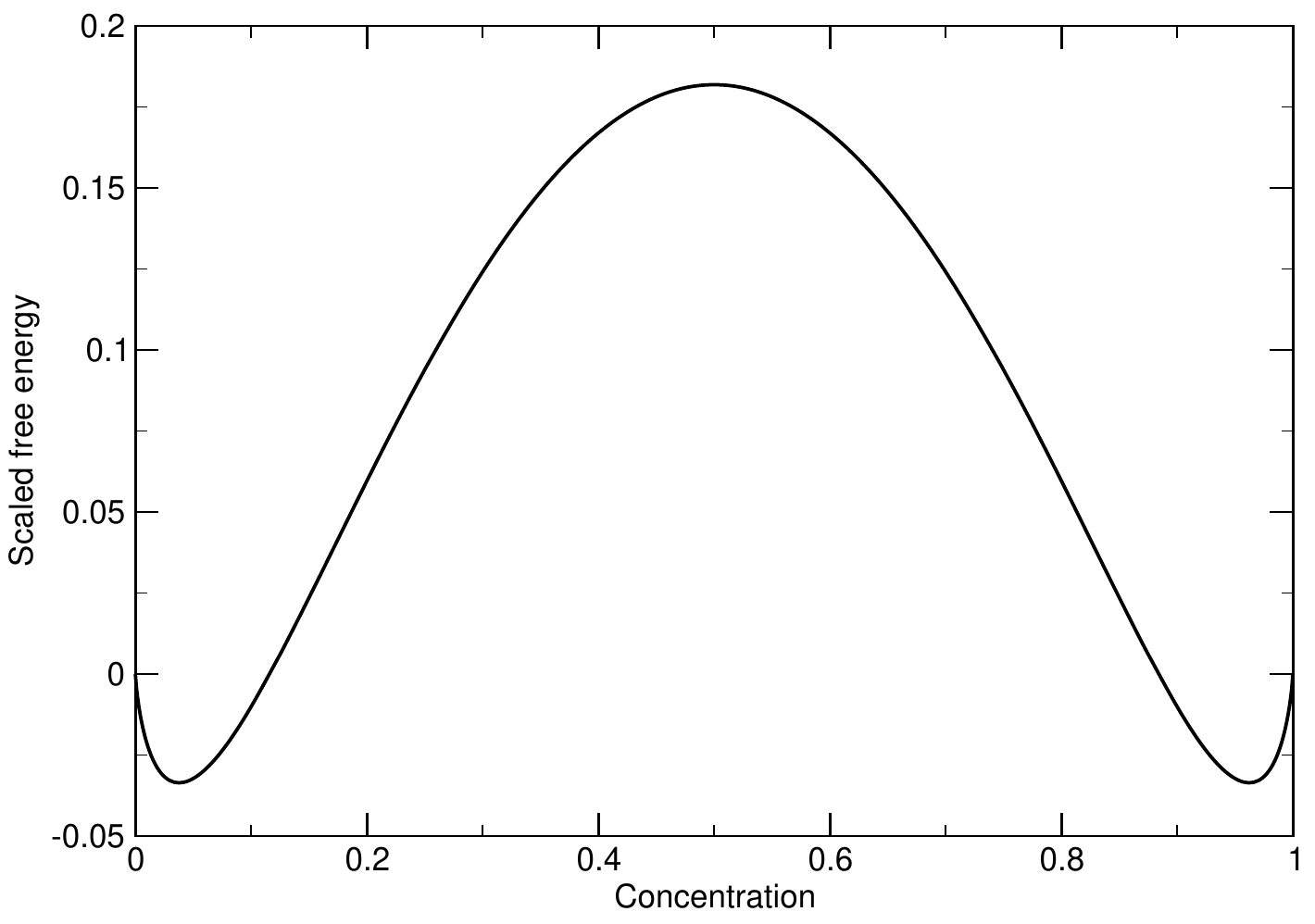}
    \includegraphics[width=0.50\textwidth]{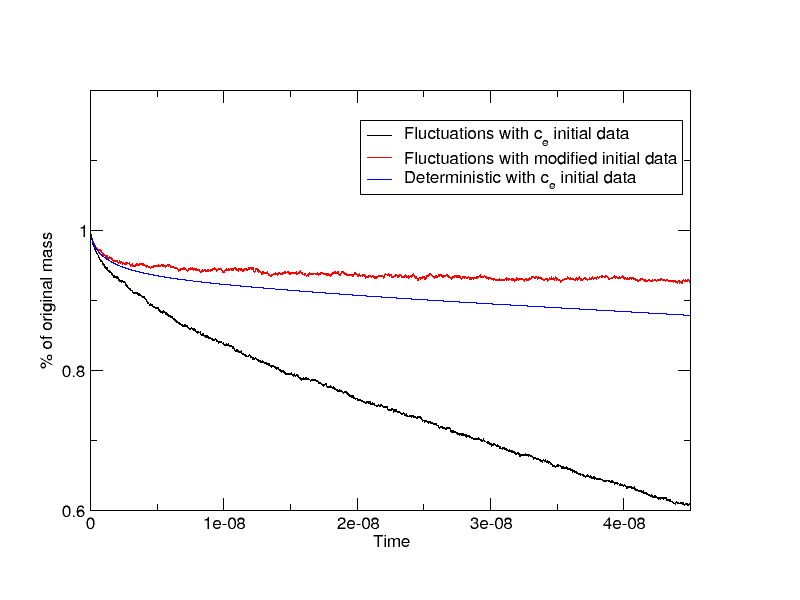}
    \caption{(Left) Scaled free energy $\mathcal{G}/\rho \kB T$ versus concentration for $\chi = 3.5$ in the homogeneous case ($\nabla c=0$). Note the asymmetry about the deterministic local minima $c_{1,e} = 0.037874$ and $c_{2,e} = 1 - c_{1,e} = 0.962126$.
    (Right) Droplet mass versus time measured using filter in Eq. (\ref{eq:filter}) and starting from discontinuous (sharp) droplet interface.  A loss of mass in the deterministic case reflects an asymmetric smoothing of the interface during relaxation from the initial state.}
    \label{fig:free_energy}
\end{figure}

For deterministic simulations, the concentrations outside and inside the droplet are given by $c_{e,1}$ and $c_{e,2} = 1-c_{e,1}$, respectively.  
When fluctuations are included, preliminary simulations showed that, with these initial conditions, the droplet would lose mass over time.  
We attribute this to the asymmetry of the free energy around $c_{e,1}$ and $c_{e,2}$ (see Fig.~\ref{fig:free_energy}) which, due to fluctuations, results in the average concentration outside the droplet being higher than $c_{e,1}$ and the average inside the droplet being lower than $c_{e,2}$.  
We found that setting the concentrations to $c_{e,1}^\dagger = 0.078 \approx 2 c_{e,1}$ outside the droplet and $c_{e,2}^\dagger = 1-c_{e,1}^\dagger = 0.922$ inside the droplet stabilized the net mass in the droplet over time (see Fig.~\ref{fig:free_energy}).

Another issue when fluctuations are included is defining the location of the droplet.  
In deterministic simulations one can identify any point with concentration $c > \frac12$ as being within the droplet.
However, in stochastic simulations care must be taken to avoid mistaking points outside the droplet with large concentration fluctuations as being inside the droplet.  
As in previous work\cite{RayPlatFHD}, for data from stochastic runs we filter concentration using
\begin{align}
    \tilde{c} = \max \left( \min \left( \frac{ c + \Delta c_f - \frac12}{2\Delta c_f} , 1 \right), 0\right)
    \label{eq:filter}
\end{align}
Taking $\Delta c_f = 1/5$ effectively sharpens the interface by a factor of 5/2 and prevents misidentifying points outside the droplet as being within it.
The center of mass position is then computed as
\begin{equation}
(x_\mathrm{cm},y_\mathrm{cm},z_\mathrm{cm}) = 
\frac{\int \int \int  (x, y, z) \times \tilde{c}(x,y,z) \; dx \, dy \,dz}{\int \int \int \tilde{c}(x,y,z) \; dx \, dy \,dz} \;\;\;.
\end{equation}
The variance of this position due to Brownian motion is used to calculate the diffusion coefficient for both spherical and sessile droplets (see eqn.~(\ref{eq:VariancePosition})). 

\subsection{Validation Tests}

 Deterministic simulations were run to measure contact angles and Laplace pressures for a variety of 2D and 3D surface droplets. 
 Tables~\ref{tab:AnglePressure2D} and \ref{tab:AnglePressure3D} list results for the measured height $h$ and center-line width $w$ and show that the measured contact angle $\theta_* = 2 \tan^{-1}(h/w)$ is in good agreement with the imposed boundary condition contact angle, $\StaticCA$. 
 Similarly, the measured Laplace pressure, $\delta p$, is in good agreement with the expected value $\gamma/R$ for 2D droplets and $2\gamma/R$ for 3D droplets. 
 In both cases the radius of curvature is measured as $R = h/(1 - \cos\theta_*)$. 
 For a 2D (cylindrical, $R = 9.79$~nm) droplet in bulk fluid (periodic boundaries) the measured Laplace pressure of 2.496~MPa is in excellent agreement with the expected value of 2.484~MPa.
 An analogous test for a 3D droplet with $R = 9.76$ nm resulted in a measured pressure of 5.01 MPa compared the theoretical value of 4.99 MPa.

\begin{table}[h]
    \centering
    \begin{tabular}{c|c|c|c|c|c|c|c}
       $\StaticCA$ & $30^\circ$ & $45^\circ$ & $60^\circ$ & $90^\circ$ & $120^\circ$ & $135^\circ$ & $150^\circ$ \\
        \hline\hline
        $h$~(nm) & 8.03 & 9.72 & 11.21 & 13.95 & 16.39 & 17.52 & 18.47 \\
        $w$~(nm) & 27.8 & 22.76 & 19.25 & 13.95 & 8.66 & 7.60 & 5.63 \\
        $\theta_*$ & $32.2^\circ$ & $46.25^\circ$ & $60.69^\circ$ & $90^\circ$ & $118.97^\circ$ & $133.09^\circ$ & $146.07^\circ$ \\
        \hline
        $\delta p$~(MPa) & 0.456 & 0.767 & 1.11  & 1.75 & 2.22 & 2.36 & 2.41 \\
        $R$~(nm) & 53.4 & 31.7 & 21.9  & 13.9 & 11.0 & 10.3 & 10.1 \\
        $\gamma/R$~(MPa) & 0.467 & 0.772 & 1.11  & 1.74 & 2.20 & 2.36 & 2.41
    \end{tabular}
    \caption{
     Results from validation tests for 2D (truncated cylinder) sessile droplets. 
     \label{tab:AnglePressure2D}
    }
\end{table}

\begin{table}[h]
    \centering
    \begin{tabular}{c|c|c|c}
       $\StaticCA$ & $45^\circ$ & $90^\circ$ &  $135^\circ$  \\
        \hline\hline
        $h$~(nm) & 7.71 & 12.38 & 16.07 \\
        $w$~(nm) & 17.61 & 12.38 & 7.25  \\
        $\theta_*$  & $47.29^\circ$ & $90^\circ$ & $133.69^\circ$  \\
        \hline
        $\delta p$~(MPa)  & 2.06 & 4.00  & 4.93 \\
        $R$~(nm) & 24.0 & 12.4 & 10.0 \\
        $2 \gamma/R$~(MPa) & 2.03 & 3.93 & 4.85 
    \end{tabular}
    \caption{
    Results from validation tests for 3D (spherical cap) sessile droplets.
    }
    \label{tab:AnglePressure3D}
\end{table}

\newpage

\subsection{Brownian Motion of Spherical Droplets}

We performed a variety of simulations to measure the diffusion coefficient of spherical droplets.
First, from 3D deterministic runs we measured the droplets' mobility and used
the Einstein relation
\[
\DiffE = \kB T \mathcal{M},
\label{eq:DiffE}
\] 
which relates the diffusion coefficient to the mobility, $\mathcal{M}$.
The mobility of a droplet is determined by applying a body force given by
\begin{equation}
F(c) = F_0 \; \frac{c-c_{e,1}}{c_{e,2}-c_{e,1}}
\label{eq:BodyForce}
\end{equation}
to the fluid within a droplet, where $F_0 = 4.0 \times 10^{7}$~N.
The simulations were run deterministically, that is, with the stochastic fluxes set to zero until a terminal velocity was reached.
The net applied force, $F_\mathrm{net}$, and the droplet's steady state (terminal) velocity, $u_\mathrm{T}$, gives the mobility from $\mathcal{M} = u_\mathrm{T}/F_\mathrm{net}$.
We note that for periodic and slip boundary conditions, the entire system is accelerated so
we account for this effect by measuring the terminal velocity in the frame of reference of the exterior fluid. That is, we take the difference between the velocity of the center of mass of the droplet and the minimum fluid velocity outside of the droplet.

For spherical droplets the Stokes estimate for the mobility is
\begin{equation}
    \mathcal{M}_\mathrm{S} = \frac{\mathcal{P}}{f_\eta \pi \eta_\mathrm{o} R}~
\end{equation}
where the viscous drag factor is\cite{HappelBrenner2012}
\begin{equation}
    f_\eta = \frac{6 + 4 \alpha}{1 + \alpha}
    \label{eq:ViscDragFactor}
\end{equation}
with $\alpha = \eta_\mathrm{o}/\eta_\mathrm{i}$ being the ratio
of the viscosities of the outer and inner fluids;
in our case $\alpha = 1$ so $f_\eta = 5$.
The finite domain correction for a cubic periodic domain is
$\mathcal{P} \approx 1 - 2.84 \,{R}/{L}$; 
for $R = 10$~nm, $L = 128$~nm this gives $\mathcal{P} \approx 0.78$.\cite{PeriodicLatticeDrag1959,dunweg1993molecular}
While the Stokes-Einstein relation, $\DiffSE = \kB T \mathcal{M}_\mathrm{S}$, is widely accepted some molecular dynamics simulations have raised issues regarding its universality.\cite{li2009BreakdownSE,Kawasaki2017BreakdownSE}

Stochastic simulations were run to measure the diffusion coefficient from the Brownian motion of the droplets. 
The center of mass position, $[x_\mathrm{cm}(t), y_\mathrm{cm}(t), z_\mathrm{cm}(t)]$, is recorded and, after skipping the first 10\% of the total run, an average diffusion coefficient is obtained by fitting the remaining data to
\begin{equation}
    \langle (x_\mathrm{cm}(t+\tau) - x_\mathrm{cm}(t))^2 \rangle = 2 \langle D_x \rangle \tau
    \label{eq:VariancePosition}
\end{equation}
with similar expressions for $\langle D_y \rangle$ and $\langle D_z \rangle$.
For spherical droplets in bulk fluid we take $\DiffB = \frac13( \langle D_x \rangle + \langle D_y \rangle + \langle D_z \rangle)$; for droplets near a surface ($y=0$ plane) we use $\DiffB = \frac12( \langle D_x \rangle + \langle D_z \rangle)$.

\begin{figure}[h]
    \centering
    \includegraphics[width=0.45\textwidth]{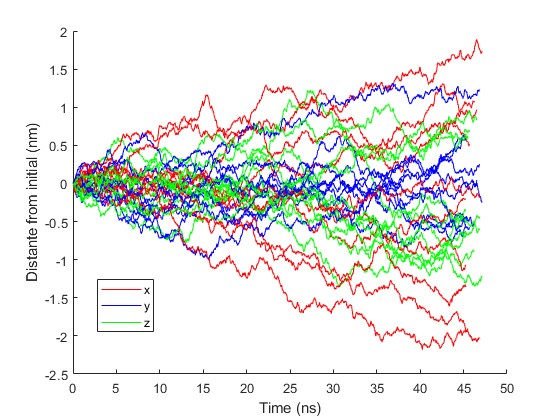}
    \includegraphics[width=0.45\textwidth]{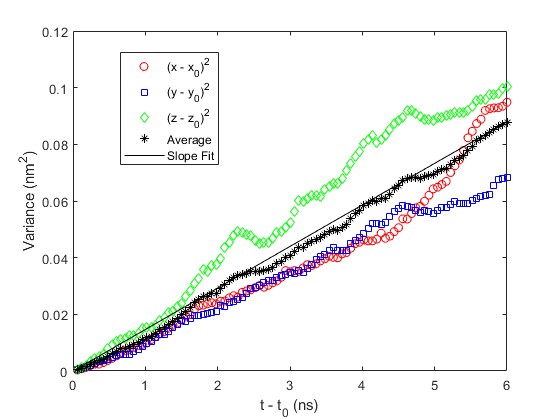}
    \caption{Spherical droplet in periodic system. (Left) Center of mass position ($x,y,z$) versus time for ensemble of 10 runs. (Right) Ensemble averaged variance of position versus time.
    }
    \label{fig:freeD}
\end{figure}

Figure~\ref{fig:freeD} shows results for a spherical droplet in bulk fluid. Statistics were collected for an ensemble of 10 runs for a free droplet of initial radius $R = 10$~nm in a cubic periodic system (i.e., no solid surfaces) of width $L = 128$~nm.
From the first column of Table~\ref{tab:MobilityDiffusion} we see that the diffusion coefficient, $\DiffE = 69.7$~nSt, obtained from the measured mobility is in good agreement with that obtained from the variance of position from Brownian motion, $\DiffB = 73.3 \pm 2.6$~nSt. 
The Stokes-Einstein value, $\DiffSE = 71.1$~nSt, is also in good agreement with these results.
It should be noted that the radii of the droplets in the deterministic and stochastic runs differed by about 1\% ($R \approx 9.7$~nm and $R \approx 9.8$~nm, respectively).

\begin{figure}[h]
    \centering
    \includegraphics[width=0.45\textwidth]{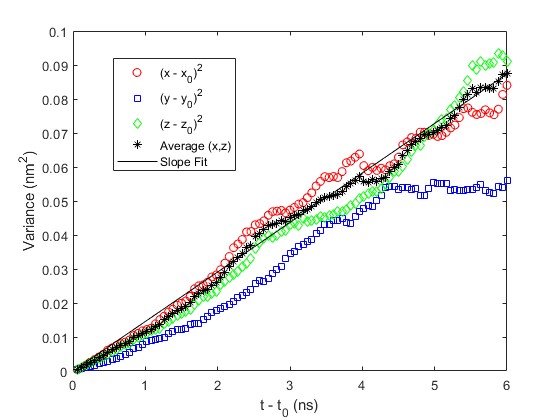}
    \includegraphics[width=0.45\textwidth]{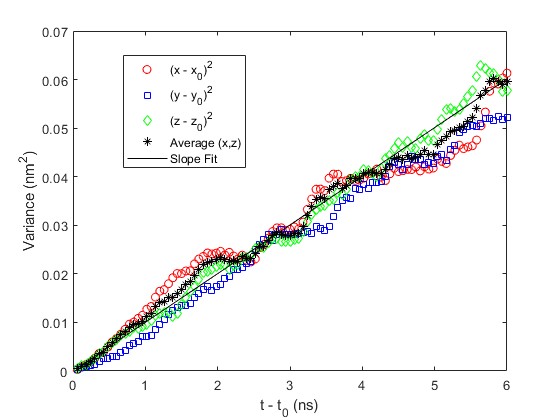}
    \caption{Variance of position versus time for a spherical droplet initially near a solid surface with slip (left) or no-slip (right) boundary conditions. Only the variance parallel to the surface is used to obtain $\DiffB$.}
    \label{fig:specialD}
\end{figure}

Figure~\ref{fig:specialD} shows results for a similar spherical droplet initialized a distance $Y$ from a solid surface ($Y = y_\mathrm{cm} = 16$~nm).
The Stokes mobility of a solid particle in an infinite domain moving parallel to a flat surface is 
\begin{align}
     \mathcal{M}_\mathrm{S}^\parallel \approx \mathcal{M}_\mathrm{S}/(1 - (3/8)(R/Y) + (9/64)(R/Y)^2 - (59/512)(R/Y)^3)
\end{align}
for a slip wall and
\begin{align}
     \mathcal{M}_\mathrm{S}^\parallel \approx \mathcal{M}_\mathrm{S} (1 - (9/16)(R/Y) + (1/8)(R/Y)^3)
\end{align}
for a no-slip wall.\cite{HappelBrenner2012}
For $R=9.65$~nm and $Y = 16$~nm we have $\mathcal{M}_\mathrm{S}^\parallel \approx 1.25 \mathcal{M}_\mathrm{S}$ for a slip wall and 
$\mathcal{M}_\mathrm{S}^\parallel \approx 0.688 \mathcal{M}_\mathrm{S}$ for a no-slip wall. 
For these cases, mobility measurements in deterministic simulations give $\mathcal{M}^\parallel \approx 1.42 \mathcal{M}$ for a slip wall and 
$\mathcal{M}^\parallel \approx 0.750 \mathcal{M}$ for a no-slip wall.
The small differences between the theoretical and measured values of mobility near a wall may be due to the finite interface thickness of the droplets, the fact that they are not solid particles, and that the periodic domain is finite.
Under the applied body force, Eq.~\ref{eq:BodyForce}, the terminal velocity of two droplets separated by a distance of $2Y = 32$nm in a $128\times256\times128$ periodic domain was measured to be identical to that of a single droplet a distance $Y = 16$nm  from a slip wall in a $128\times128\times128$ domain.  Consequently the diffusion coefficient $D_E$ for the droplets is the same in these two cases.

Table~\ref{tab:MobilityDiffusion} does show that for a droplet near a slip wall, as expected, the Einstein diffusion coefficient, $\DiffE = 98.81$~nSt, is larger than that of a droplet in the bulk fluid case and in reasonable agreement ($\lesssim 10\%$ difference) with the Stokes-Einstein prediction of $\DiffSE = 89.53$~nSt. 
However, from the measurements of Brownian motion we find that $\DiffB = 72.8 \pm 3.3$~nSt is significantly lower than the deterministic measurement
and nearly the same as for a droplet in bulk fluid.
A simulation of two droplets in a $128\times256\times128$ periodic domain for the stochastic case gave an average diffusion coefficient of $\DiffB =57.6$~nSt.
That is, in the stochastic case, two droplets separated by a distance of $2Y = 32$nm is not the same as a single droplet at a distance $Y = 16$nm near a slip wall because the stochastic terms break the symmetry associated with the slip wall.
On the other hand, near a no-slip wall the mobility decreases relative to the bulk case, which is confirmed with the values $\DiffE = 52.30$~nSt, $\DiffSE = 49.27$~nSt, and $\DiffB = 50.1 \pm 3.3$~nSt being in good agreement (see Table~\ref{tab:MobilityDiffusion}).

\begin{figure}[h]
    \centering
      \includegraphics[width=0.49\textwidth]{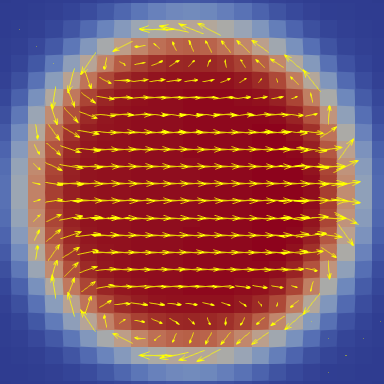}\\
    \includegraphics[width=0.49\textwidth]{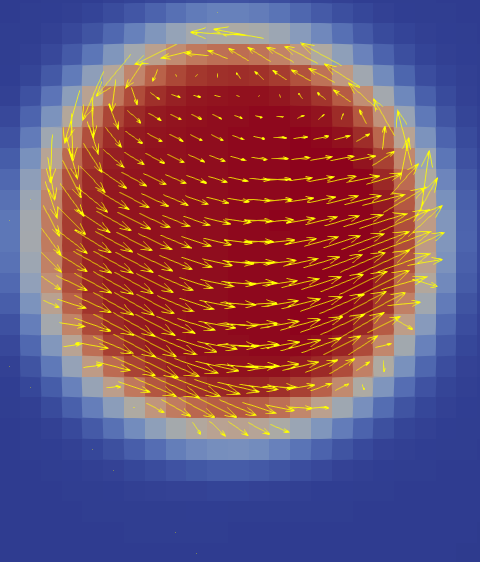}
   \includegraphics[width=0.49\textwidth]{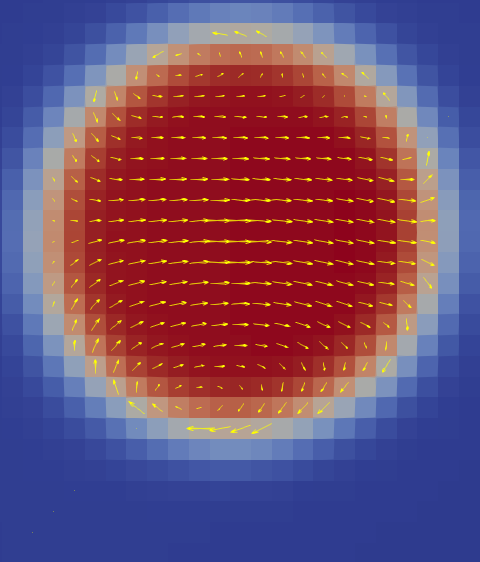}
    \caption{Flow inside a spherical droplet ($R \approx 10$~nm) moving at terminal velocity due to an applied force in the $+x$ direction. (Top) Droplet in bulk fluid (periodic boundaries). Droplet is near a solid surface with slip  (lower left) and no-slip (lower right) boundary conditions. The surface is 16~nm from the droplet's center and located at the bottom edge of the image.}
    \label{fig:sphereVel}
\end{figure}

The flow profiles for the fluid within the spherical droplets moving at terminal velocity are shown in Figure~\ref{fig:sphereVel}; for visual clarity the flow outside the droplets is not plotted.
This data is taken from the deterministic simulations in which an external force was applied to measure the mobility.
The applied force is in the $+x$ direction and the velocity in the figure is shifted to a frame of reference in which the droplet is stationary (i.e., shifted by the center of mass velocity).
Note that for a spherical droplet moving parallel to a slip wall the flow is noticeably asymmetric and counter-clockwise within the droplet. By comparison, for the droplets in bulk fluid or droplets moving parallel to a no-slip wall the flow is counter-clockwise above the mid-line and clockwise below it. 
The droplet in the bulk and the droplet near the no-slip wall have two recirculation zones while the droplet near the slip wall has only one.

\begin{table}[h]
    \centering
    \begin{tabular}{c||c|c|c||c|c|c||c|c|c}
        ~ &\multicolumn{3}{c||}{Sphere} & \multicolumn{3}{c||}{Sessile (Slip)} & \multicolumn{3}{c}{Sessile (No-Slip)} \\
        ~ & Bulk & Slip & No-slip & $45^\circ$ & $90^\circ$ & $135^\circ$ & $45^\circ$ & $90^\circ$ & $135^\circ$  \\
        \hline
        $\DiffE$~(nSt) & 69.7 & 98.81 & 52.3 & 128.1 & 131.1 & 122.0  & 10.97 & 17.28 & 22.88 \\
        $\DiffB$~(nSt) & 73.3 & 72.8 & 50.1 & 116.0 & 111.1  & 97.3 & 12.2 & 24.1 & 33.6 \\
        $\sigma_\mathrm{B}$~(nSt)  & 2.6 & 3.3 & 3.2 & 7.0 & 4.9 & 4.2 & 1.0 & 1.7 & 2.5
    \end{tabular}
    \caption{The diffusion coefficient, $\DiffE = \kB T \mathcal{M}$, from the Einstein relation and $\DiffB$ computed from the variance of position (see eqn.~(\ref{eq:VariancePosition})); the standard error, $\sigma_\mathrm{B}$, is for the latter. The first three columns are for a spherical drop, either in bulk fluid or near a slip or no-slip surface. Remaining columns are for sessile droplets with various contact angles and fluid velocity boundary conditions on the surface. }
    \label{tab:MobilityDiffusion}
\end{table}

\clearpage

\subsection{Brownian Motion of Sessile Droplets}

We also studied the dynamics of sessile (surface) droplets, with a variety of contact angles, moving on a flat, solid surface, with both slip and no-slip boundary conditions for the fluid. The initial radii were selected so that the droplets had the same mass as the free spherical droplet considered above.  Specifically, we used $R = $25.8nm, 12.6nm, and 10.2nm for contact angles of $45^\circ$,  $90^\circ$, and  $135^\circ$, respectively. 
The mobility and diffusion coefficients were measured as described in the previous section; note that only the variance in position parallel to the surface was used in computing $\DiffB$. 
As a special validation test we verified that the deterministically measured mobility of a $90^\circ$ sessile droplet on a slip wall in a domain with $N_y = 64$ matched that of a spherical droplet in a periodic domain with $N_y = 128$ (with $N_x = N_z = 128$ in both cases).

\begin{figure}[h]
    \centering
    \includegraphics[width=0.45\textwidth]{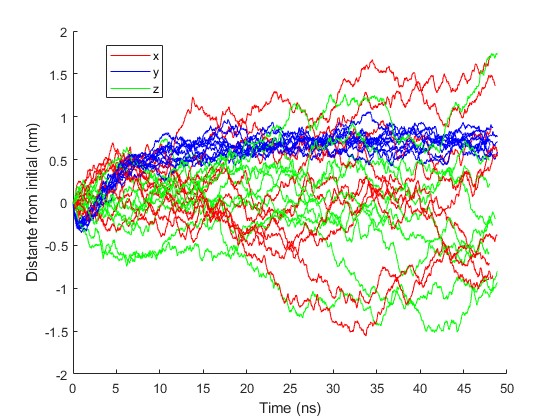}
    \includegraphics[width=0.45\textwidth]{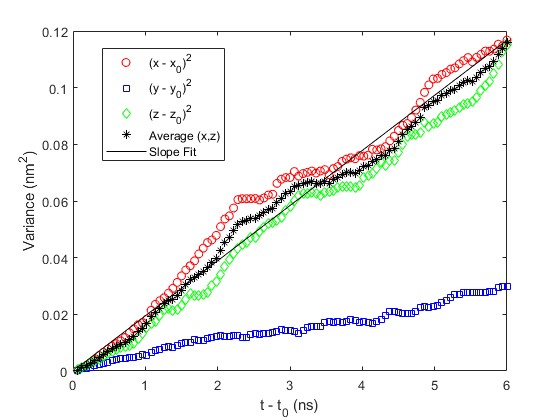} \\
    \includegraphics[width=0.45\textwidth]{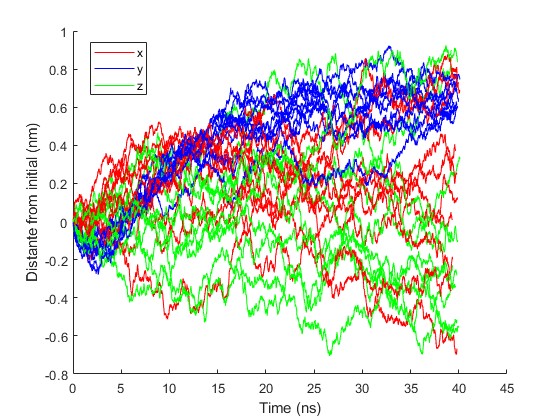}
    \includegraphics[width=0.45\textwidth]{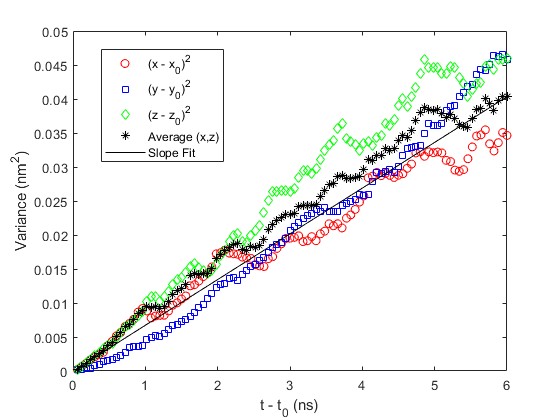}
    \caption{Sessile droplets with $\StaticCA = 135^\circ$ contact angle on (top) slip surface or (bottom) no-slip surface.
    (Left) Center of mass position ($x,y,z$) versus time for the ensemble of 10 runs. (Right) Variance of position versus time.
    }
    \label{fig:135D}
\end{figure}

\begin{figure}[h]
    \centering
    \includegraphics[width=0.45\textwidth]{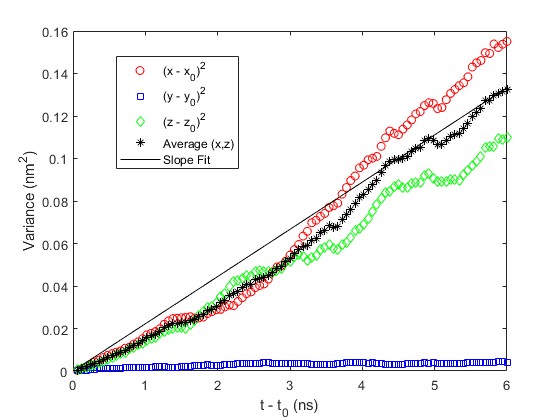}
    \includegraphics[width=0.45\textwidth]{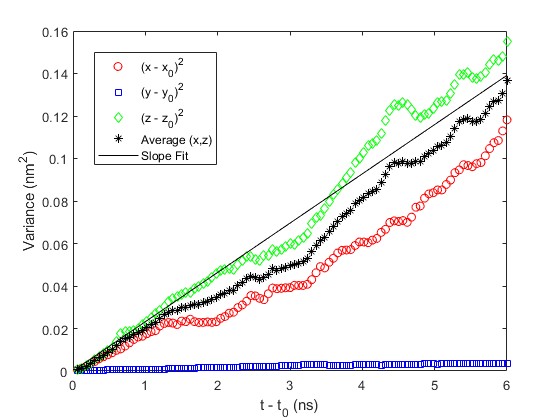} \\
    \includegraphics[width=0.45\textwidth]{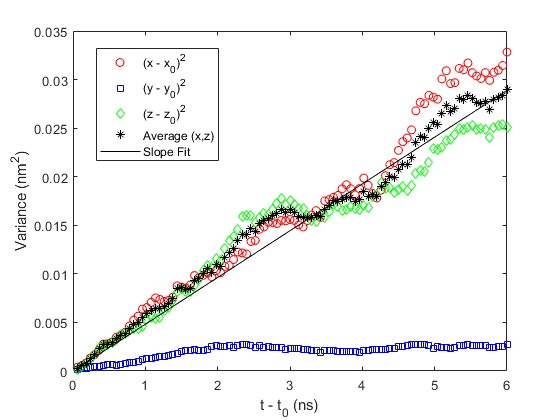}
    \includegraphics[width=0.45\textwidth]{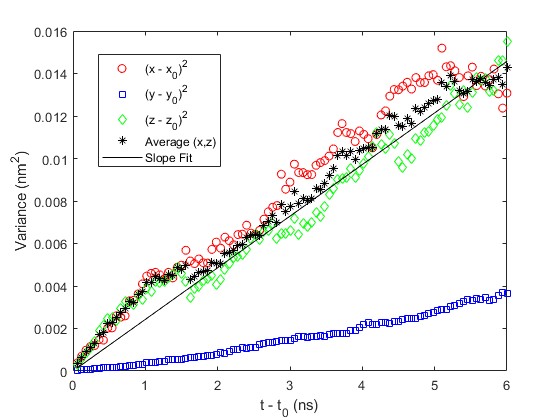}
    \caption{Variance of position versus time for sessile droplets with (left) $\StaticCA = 90^\circ$ or (right) $\StaticCA = 45^\circ$ contact angle on (top) slip surface or (bottom) no-slip surface.}
    \label{fig:90_45D}
\end{figure}

Figure~\ref{fig:135D} shows results for the Brownian motion of sessile droplets on a hydrophobic surface ($\StaticCA = 135^\circ$).
Results for sessile droplets on a neutral surface ($\StaticCA = 90^\circ$) and a hydrophilic surface ($\StaticCA = 45^\circ$) are shown in Figure~\ref{fig:90_45D}.
Though only the variance in the horizontal position is used in calculating $\DiffB$ we see in these figures that there is significant variance in $y_\mathrm{cm}$, especially for no-slip velocity boundary conditions.
Note that in the molecular dynamics study by Niu \textit{et al.}\cite{DropDiffusionMD_2018} they found that the variance of horizontal position (see Eq.~\ref{eq:VariancePosition}) went as $\tau^\mathfrak{g}$ with $\mathfrak{g} < 1$ at very short (sub-picosecond) timescales and $\mathfrak{g} > 1$ at intermediate (picosecond) timescales. At nanosecond timescales they reported $\mathfrak{g} \approx 1$, which is consistent with our findings, given the statistical errors in the data.

From Table~\ref{tab:MobilityDiffusion} we see that for the hydrophilic cases (contact angle $\StaticCA = 45^\circ$) we have $\DiffE \approx \DiffB$ for both slip and no-slip  surfaces.
In the other cases we find $\DiffE > \DiffB$ for a slip surface while $\DiffE < \DiffB$ for a no-slip surface.
Specifically, on a slip surface $\DiffE / \DiffB = 1.10 \pm 0.07$ for $45^\circ$; $\DiffE / \DiffB = 1.18 \pm 0.05$ for $90^\circ$; $\DiffE / \DiffB = 1.25 \pm 0.05$ for $135^\circ$ so this ratio increases with increasing contact angle.
To validate that the discrepancy at the higher contact angle was not an artifact caused by the strength of the forcing we performed an additional test in which the forcing was reduced by a factor of 4.  
Although this weaker forcing reduced the deformation of the droplet, the mobility was unchanged.
By comparison, on a no-slip surface $\DiffE / \DiffB = 0.90 \pm 0.07$ for $45^\circ$; $\DiffE / \DiffB = 0.72 \pm 0.05$ for $90^\circ$; $\DiffE / \DiffB = 0.68 \pm 0.05$ for $135^\circ$ so this ratio decreases with increasing contact angle.
On a no-slip surface both $\DiffE$ and $\DiffB$ increase with increasing contact angle and, not surprisingly, are significantly smaller than for droplets on a slip surface.
The molecular dynamics results of Li \textit{et al.}\cite{DropDiffusionMD_2016}
suggested that $\DiffB \propto 1/R \sin(\StaticCA) (1 + \cos(\StaticCA))^2$ and we find that to be approximately true ($\pm 10\%$) for no-slip surfaces but not in the case of slip boundary conditions.
The many-body dissipative particle dynamics simulations of sessile droplets by Chang \textit{et al.}\cite{chang2016wetting} found that $\DiffB$ decreases exponentially with contact area, $\pi R^2 \sin^2(\StaticCA)$, and our results for no-slip surfaces are also consistent with their findings.

\begin{figure}[h]
    \centering
    \includegraphics[width=\textwidth]{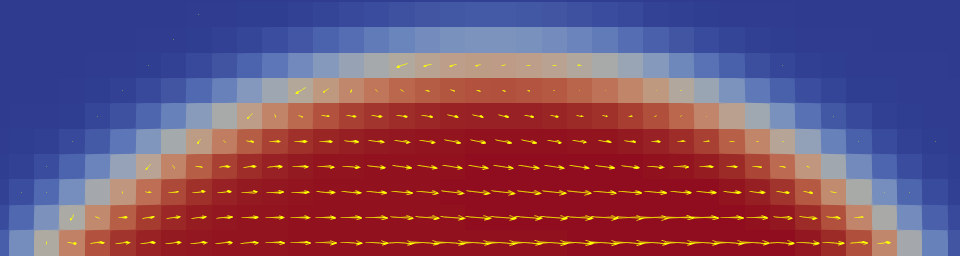}\\
    \includegraphics[width=0.5\textwidth]{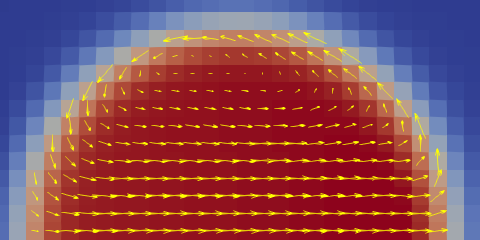}
    \includegraphics[width=0.4\textwidth]{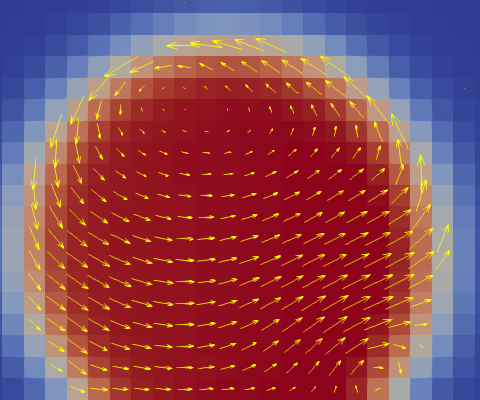}
    \caption{Flow inside droplets moving on a slip surface under an applied force (in the $+x$ direction) for boundary condition contact angles:  $\StaticCA =$ $45^\circ$ (top); $90^\circ$ (middle); $135^\circ$ (bottom).}
    \label{fig:slipVel}
\end{figure}

\begin{figure}[h]
    \centering
    \includegraphics[width=\textwidth]{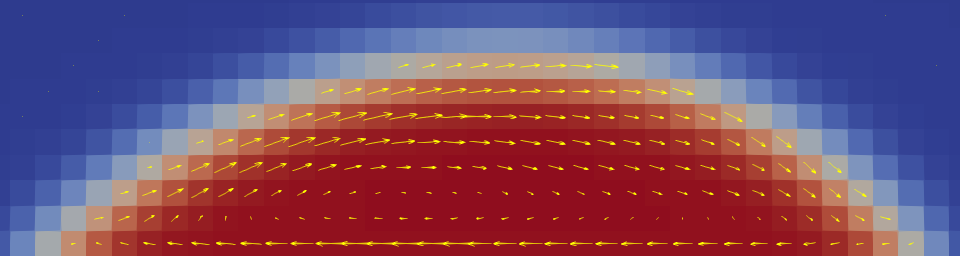}\\
    \includegraphics[width=0.5\textwidth]{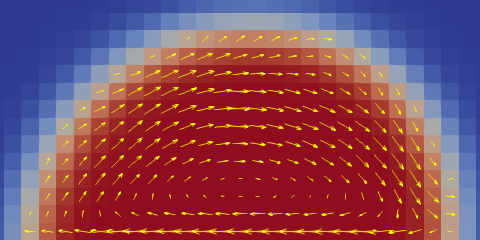}
    \includegraphics[width=0.4\textwidth]{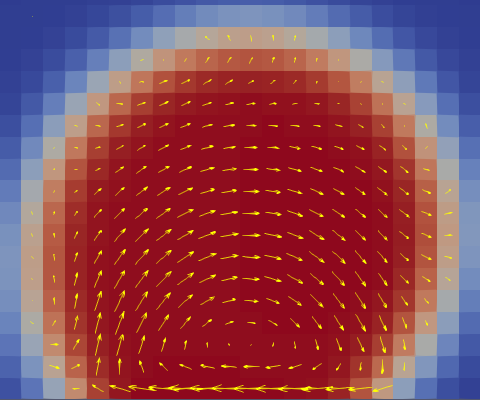}
    \caption{Flow inside droplets moving on a no-slip surface under an applied force (in the $+x$ direction) for boundary condition contact angles:  $\StaticCA =$ $45^\circ$ (top); $90^\circ$ (middle); $135^\circ$ (bottom).}
    \label{fig:noslipVel}
\end{figure}

The flow profiles for sessile droplets moving at terminal velocity are shown in Figure~\ref{fig:slipVel} for a slip surface and Figure~\ref{fig:noslipVel} for a no-slip surface.
Since the applied force is in the $+x$ direction and these snapshots are in the frame of reference in which the droplet is stationary it is not surprising that the flow is counter-clockwise in drops on a slip surface and clockwise in drops on a no-slip surface.
For droplets near a slip wall, the center of recirculation is near the top of the droplet and for a contact angle of $135^\circ$ there is a secondary recirculation region near the slip wall at the leading edge of the droplet.  
For droplets on a no-slip surface the center of the recirculation is near the wall.  
The shape of the sessile droplets is somewhat distorted and becomes asymmetric, particularly for the no-slip cases, which is consistent with the observations of Chang \textit{et al.}\cite{chang2016wetting}.

\clearpage

\section{Concluding Remarks}

We employ a Cahn-Hillard formulation coupled with incompressible fluctuating hydrodynamics to simulate the mesoscopic dynamics of phase-separated droplets at the nanoscale. We consider both spherical fluid droplets immersed in a second fluid and sessile droplets with a specified contact angle incorporated as a boundary condition. Deterministic simulations with an applied body force allow for the measurement of droplet mobility and the calculation of a diffusion coefficient using the Einstein relation. Stochastic simulations provide an independent method for obtaining the diffusion coefficient by fitting the variance of droplet position over time.

In our simulations we find that the diffusion coefficient, $\DiffE$, given by the Einstein relation (Eq.~\ref{eq:DiffE}) is not always in agreement with the diffusion coefficient, $\DiffB$, given by the variance of position in Brownian motion (Eq.~\ref{eq:VariancePosition}). 
From Table~\ref{tab:MobilityDiffusion} we see that $\DiffE \approx \DiffB$ for 1) spherical droplets in a periodic domain; 2) spherical droplets near a no-slip wall; 3) sessile droplets on a hydrophilic surface ($\StaticCA = 45^\circ$) for both slip and no-slip conditions.
However $D_E > D_B$ for 1) spherical droplets near a slip wall; 2) sessile droplets on a slip wall that is neutral ($\StaticCA = 90^\circ$) or hydrophobic ($\StaticCA = 135^\circ$).
Finally, $D_E < D_B$ for sessile droplets on a no-slip wall that is neutral ($\StaticCA = 90^\circ$) or hydrophobic ($\StaticCA = 135^\circ$).

Fluctuating hydrodynamics (FHD) provides a useful, complementary simulation approach for the study of fluid phenomena at the mesoscale where thermal fluctuations are significant. In contrast with molecular dynamics where the fluid and solid interface properties are prescribed by the intermolecular forces in FHD one may independently set viscosity, surface tension, contact angle, etc.
There are various avenues for future work.
Qin \textit{et al.}~\cite{DropletsNerstCH_2023} used a Poisson–Nernst–Planck –Navier–Stokes–Cahn–Hillard (PNP–NS–CH) model for an electrically charged droplet suspended in a viscous fluid under an external electric field. Fluctuations can be incorporated into this deterministic model using the methodology of Peraud, \textit{et al.}.~\cite{peraud2016}
Another potetial application is the study of surfactant phenomena (e.g., Marangoni effect) in nanoscale droplets.\cite{Teigen2011surfactant,Zong2020surfactant,Arbabi2023surfactant}

\section*{Acknowledgements}
The authors thank Profs. Matthew Borg, James Sprittles and Duncan Lockerby for fruitful discussions.
This work was supported by the U.S. Department of Energy, Office of Science, Office of Advanced Scientific Computing Research, Applied Mathematics Program under contract No. DE-AC02-05CH11231.

\bibliography{ContactAngle}

\begin{thebibliography}{52}%
\makeatletter
\providecommand \@ifxundefined [1]{%
 \@ifx{#1\undefined}
}%
\providecommand \@ifnum [1]{%
 \ifnum #1\expandafter \@firstoftwo
 \else \expandafter \@secondoftwo
 \fi
}%
\providecommand \@ifx [1]{%
 \ifx #1\expandafter \@firstoftwo
 \else \expandafter \@secondoftwo
 \fi
}%
\providecommand \natexlab [1]{#1}%
\providecommand \enquote  [1]{``#1''}%
\providecommand \bibnamefont  [1]{#1}%
\providecommand \bibfnamefont [1]{#1}%
\providecommand \citenamefont [1]{#1}%
\providecommand \href@noop [0]{\@secondoftwo}%
\providecommand \href [0]{\begingroup \@sanitize@url \@href}%
\providecommand \@href[1]{\@@startlink{#1}\@@href}%
\providecommand \@@href[1]{\endgroup#1\@@endlink}%
\providecommand \@sanitize@url [0]{\catcode `\\12\catcode `\$12\catcode `\&12\catcode `\#12\catcode `\^12\catcode `\_12\catcode `\%12\relax}%
\providecommand \@@startlink[1]{}%
\providecommand \@@endlink[0]{}%
\providecommand \url  [0]{\begingroup\@sanitize@url \@url }%
\providecommand \@url [1]{\endgroup\@href {#1}{\urlprefix }}%
\providecommand \urlprefix  [0]{URL }%
\providecommand \Eprint [0]{\href }%
\providecommand \doibase [0]{http://dx.doi.org/}%
\providecommand \selectlanguage [0]{\@gobble}%
\providecommand \bibinfo  [0]{\@secondoftwo}%
\providecommand \bibfield  [0]{\@secondoftwo}%
\providecommand \translation [1]{[#1]}%
\providecommand \BibitemOpen [0]{}%
\providecommand \bibitemStop [0]{}%
\providecommand \bibitemNoStop [0]{.\EOS\space}%
\providecommand \EOS [0]{\spacefactor3000\relax}%
\providecommand \BibitemShut  [1]{\csname bibitem#1\endcsname}%
\let\auto@bib@innerbib\@empty
\bibitem [{\citenamefont {Bonn}\ \emph {et~al.}(2009)\citenamefont {Bonn}, \citenamefont {Eggers}, \citenamefont {Indekeu}, \citenamefont {Meunier},\ and\ \citenamefont {Rolley}}]{bonn2009wetting}%
  \BibitemOpen
  \bibfield  {author} {\bibinfo {author} {\bibfnamefont {D.}~\bibnamefont {Bonn}}, \bibinfo {author} {\bibfnamefont {J.}~\bibnamefont {Eggers}}, \bibinfo {author} {\bibfnamefont {J.}~\bibnamefont {Indekeu}}, \bibinfo {author} {\bibfnamefont {J.}~\bibnamefont {Meunier}}, \ and\ \bibinfo {author} {\bibfnamefont {E.}~\bibnamefont {Rolley}},\ }\bibfield  {title} {\enquote {\bibinfo {title} {Wetting and spreading},}\ }\href@noop {} {\bibfield  {journal} {\bibinfo  {journal} {Reviews of modern physics}\ }\textbf {\bibinfo {volume} {81}},\ \bibinfo {pages} {739--805} (\bibinfo {year} {2009})}\BibitemShut {NoStop}%
\bibitem [{\citenamefont {Lohse}\ and\ \citenamefont {Zhang}(2015)}]{lohse2015surface}%
  \BibitemOpen
  \bibfield  {author} {\bibinfo {author} {\bibfnamefont {D.}~\bibnamefont {Lohse}}\ and\ \bibinfo {author} {\bibfnamefont {X.}~\bibnamefont {Zhang}},\ }\bibfield  {title} {\enquote {\bibinfo {title} {Surface nanobubbles and nanodroplets},}\ }\href@noop {} {\bibfield  {journal} {\bibinfo  {journal} {Reviews of modern physics}\ }\textbf {\bibinfo {volume} {87}},\ \bibinfo {pages} {981--1035} (\bibinfo {year} {2015})}\BibitemShut {NoStop}%
\bibitem [{\citenamefont {Qian}, \citenamefont {Arends},\ and\ \citenamefont {Zhang}(2019)}]{Qian2019nanodroplets}%
  \BibitemOpen
  \bibfield  {author} {\bibinfo {author} {\bibfnamefont {J.}~\bibnamefont {Qian}}, \bibinfo {author} {\bibfnamefont {G.~F.}\ \bibnamefont {Arends}}, \ and\ \bibinfo {author} {\bibfnamefont {X.}~\bibnamefont {Zhang}},\ }\bibfield  {title} {\enquote {\bibinfo {title} {Surface nanodroplets: formation, dissolution, and applications},}\ }\href@noop {} {\bibfield  {journal} {\bibinfo  {journal} {Langmuir}\ }\textbf {\bibinfo {volume} {35}},\ \bibinfo {pages} {12583--12596} (\bibinfo {year} {2019})}\BibitemShut {NoStop}%
\bibitem [{\citenamefont {Sun}\ \emph {et~al.}(2023)\citenamefont {Sun}, \citenamefont {Zeng}, \citenamefont {Deng}, \citenamefont {Zhang},\ and\ \citenamefont {Zhang}}]{sun2023droplet}%
  \BibitemOpen
  \bibfield  {author} {\bibinfo {author} {\bibfnamefont {Z.}~\bibnamefont {Sun}}, \bibinfo {author} {\bibfnamefont {X.}~\bibnamefont {Zeng}}, \bibinfo {author} {\bibfnamefont {X.}~\bibnamefont {Deng}}, \bibinfo {author} {\bibfnamefont {X.}~\bibnamefont {Zhang}}, \ and\ \bibinfo {author} {\bibfnamefont {Y.}~\bibnamefont {Zhang}},\ }\bibfield  {title} {\enquote {\bibinfo {title} {Droplet interface in additive manufacturing: from process to application},}\ }\href@noop {} {\bibfield  {journal} {\bibinfo  {journal} {Droplet}\ }\textbf {\bibinfo {volume} {2}},\ \bibinfo {pages} {e57} (\bibinfo {year} {2023})}\BibitemShut {NoStop}%
\bibitem [{\citenamefont {Perumanath}\ \emph {et~al.}(2020)\citenamefont {Perumanath}, \citenamefont {Borg}, \citenamefont {Sprittles},\ and\ \citenamefont {Enright}}]{DropJumping2020}%
  \BibitemOpen
  \bibfield  {author} {\bibinfo {author} {\bibfnamefont {S.}~\bibnamefont {Perumanath}}, \bibinfo {author} {\bibfnamefont {M.~K.}\ \bibnamefont {Borg}}, \bibinfo {author} {\bibfnamefont {J.~E.}\ \bibnamefont {Sprittles}}, \ and\ \bibinfo {author} {\bibfnamefont {R.}~\bibnamefont {Enright}},\ }\bibfield  {title} {\enquote {\bibinfo {title} {Molecular physics of jumping nanodroplets},}\ }\href@noop {} {\bibfield  {journal} {\bibinfo  {journal} {Nanoscale}\ }\textbf {\bibinfo {volume} {12}},\ \bibinfo {pages} {20631--20637} (\bibinfo {year} {2020})}\BibitemShut {NoStop}%
\bibitem [{\citenamefont {Bureš}\ and\ \citenamefont {Sato}(2022)}]{DropBoiling_2022}%
  \BibitemOpen
  \bibfield  {author} {\bibinfo {author} {\bibfnamefont {L.}~\bibnamefont {Bureš}}\ and\ \bibinfo {author} {\bibfnamefont {Y.}~\bibnamefont {Sato}},\ }\bibfield  {title} {\enquote {\bibinfo {title} {Comprehensive simulations of boiling with a resolved microlayer: validation and sensitivity study},}\ }\href {\doibase 10.1017/jfm.2021.1108} {\bibfield  {journal} {\bibinfo  {journal} {Journal of Fluid Mechanics}\ }\textbf {\bibinfo {volume} {933}},\ \bibinfo {pages} {A54} (\bibinfo {year} {2022})}\BibitemShut {NoStop}%
\bibitem [{\citenamefont {Zhu}\ and\ \citenamefont {Fang}(2013)}]{zhu2013analytical}%
  \BibitemOpen
  \bibfield  {author} {\bibinfo {author} {\bibfnamefont {Y.}~\bibnamefont {Zhu}}\ and\ \bibinfo {author} {\bibfnamefont {Q.}~\bibnamefont {Fang}},\ }\bibfield  {title} {\enquote {\bibinfo {title} {Analytical detection techniques for droplet microfluidics—a review},}\ }\href@noop {} {\bibfield  {journal} {\bibinfo  {journal} {Analytica chimica acta}\ }\textbf {\bibinfo {volume} {787}},\ \bibinfo {pages} {24--35} (\bibinfo {year} {2013})}\BibitemShut {NoStop}%
\bibitem [{\citenamefont {Qiao}\ \emph {et~al.}(2014)\citenamefont {Qiao}, \citenamefont {Zhang}, \citenamefont {Yen}, \citenamefont {Ku}, \citenamefont {Song}, \citenamefont {Lian},\ and\ \citenamefont {Lo}}]{qiao2014oil}%
  \BibitemOpen
  \bibfield  {author} {\bibinfo {author} {\bibfnamefont {W.}~\bibnamefont {Qiao}}, \bibinfo {author} {\bibfnamefont {T.}~\bibnamefont {Zhang}}, \bibinfo {author} {\bibfnamefont {T.}~\bibnamefont {Yen}}, \bibinfo {author} {\bibfnamefont {T.-H.}\ \bibnamefont {Ku}}, \bibinfo {author} {\bibfnamefont {J.}~\bibnamefont {Song}}, \bibinfo {author} {\bibfnamefont {I.}~\bibnamefont {Lian}}, \ and\ \bibinfo {author} {\bibfnamefont {Y.-H.}\ \bibnamefont {Lo}},\ }\bibfield  {title} {\enquote {\bibinfo {title} {Oil-encapsulated nanodroplet array for bio-molecular detection},}\ }\href@noop {} {\bibfield  {journal} {\bibinfo  {journal} {Annals of biomedical engineering}\ }\textbf {\bibinfo {volume} {42}},\ \bibinfo {pages} {1932--1941} (\bibinfo {year} {2014})}\BibitemShut {NoStop}%
\bibitem [{\citenamefont {Oh}, \citenamefont {Han},\ and\ \citenamefont {Kim}(2015)}]{HydrogenBubble2015}%
  \BibitemOpen
  \bibfield  {author} {\bibinfo {author} {\bibfnamefont {S.~H.}\ \bibnamefont {Oh}}, \bibinfo {author} {\bibfnamefont {J.~G.}\ \bibnamefont {Han}}, \ and\ \bibinfo {author} {\bibfnamefont {J.-M.}\ \bibnamefont {Kim}},\ }\bibfield  {title} {\enquote {\bibinfo {title} {Long-term stability of hydrogen nanobubble fuel},}\ }\href@noop {} {\bibfield  {journal} {\bibinfo  {journal} {Fuel}\ }\textbf {\bibinfo {volume} {158}},\ \bibinfo {pages} {399--404} (\bibinfo {year} {2015})}\BibitemShut {NoStop}%
\bibitem [{\citenamefont {de~Ruijter}, \citenamefont {Blake},\ and\ \citenamefont {De~Coninck}(1999)}]{WettingMD_1999}%
  \BibitemOpen
  \bibfield  {author} {\bibinfo {author} {\bibfnamefont {M.~J.}\ \bibnamefont {de~Ruijter}}, \bibinfo {author} {\bibfnamefont {T.~D.}\ \bibnamefont {Blake}}, \ and\ \bibinfo {author} {\bibfnamefont {J.}~\bibnamefont {De~Coninck}},\ }\bibfield  {title} {\enquote {\bibinfo {title} {Dynamic wetting studied by molecular modeling simulations of droplet spreading},}\ }\href {\doibase 10.1021/la990171l} {\bibfield  {journal} {\bibinfo  {journal} {Langmuir}\ }\textbf {\bibinfo {volume} {15}},\ \bibinfo {pages} {7836--7847} (\bibinfo {year} {1999})}\BibitemShut {NoStop}%
\bibitem [{\citenamefont {Bertrand}, \citenamefont {Blake},\ and\ \citenamefont {Coninck}(2009)}]{WettingMD_2009}%
  \BibitemOpen
  \bibfield  {author} {\bibinfo {author} {\bibfnamefont {E.}~\bibnamefont {Bertrand}}, \bibinfo {author} {\bibfnamefont {T.~D.}\ \bibnamefont {Blake}}, \ and\ \bibinfo {author} {\bibfnamefont {J.~D.}\ \bibnamefont {Coninck}},\ }\bibfield  {title} {\enquote {\bibinfo {title} {Influence of solid–liquid interactions on dynamic wetting: a molecular dynamics study},}\ }\href {\doibase 10.1088/0953-8984/21/46/464124} {\bibfield  {journal} {\bibinfo  {journal} {Journal of Physics: Condensed Matter}\ }\textbf {\bibinfo {volume} {21}},\ \bibinfo {pages} {464124} (\bibinfo {year} {2009})}\BibitemShut {NoStop}%
\bibitem [{\citenamefont {Li}, \citenamefont {Huang},\ and\ \citenamefont {Li}(2016)}]{DropDiffusionMD_2016}%
  \BibitemOpen
  \bibfield  {author} {\bibinfo {author} {\bibfnamefont {C.}~\bibnamefont {Li}}, \bibinfo {author} {\bibfnamefont {J.}~\bibnamefont {Huang}}, \ and\ \bibinfo {author} {\bibfnamefont {Z.}~\bibnamefont {Li}},\ }\bibfield  {title} {\enquote {\bibinfo {title} {A relation for nanodroplet diffusion on smooth surfaces},}\ }\href@noop {} {\bibfield  {journal} {\bibinfo  {journal} {Scientific Reports}\ }\textbf {\bibinfo {volume} {6}},\ \bibinfo {pages} {26488} (\bibinfo {year} {2016})}\BibitemShut {NoStop}%
\bibitem [{\citenamefont {Niu}, \citenamefont {Huang},\ and\ \citenamefont {Chen}(2018)}]{DropDiffusionMD_2018}%
  \BibitemOpen
  \bibfield  {author} {\bibinfo {author} {\bibfnamefont {Z.-X.}\ \bibnamefont {Niu}}, \bibinfo {author} {\bibfnamefont {T.}~\bibnamefont {Huang}}, \ and\ \bibinfo {author} {\bibfnamefont {Y.}~\bibnamefont {Chen}},\ }\bibfield  {title} {\enquote {\bibinfo {title} {Molecular dynamics study of nanodroplet diffusion on smooth solid surfaces},}\ }\href@noop {} {\bibfield  {journal} {\bibinfo  {journal} {Frontiers of Physics}\ }\textbf {\bibinfo {volume} {13}},\ \bibinfo {pages} {1--6} (\bibinfo {year} {2018})}\BibitemShut {NoStop}%
\bibitem [{\citenamefont {Zhao}\ \emph {et~al.}(2022{\natexlab{a}})\citenamefont {Zhao}, \citenamefont {Zhao}, \citenamefont {Li}, \citenamefont {Yang}, \citenamefont {Chen},\ and\ \citenamefont {Xu}}]{DropMergeMD_2022}%
  \BibitemOpen
  \bibfield  {author} {\bibinfo {author} {\bibfnamefont {M.}~\bibnamefont {Zhao}}, \bibinfo {author} {\bibfnamefont {Y.}~\bibnamefont {Zhao}}, \bibinfo {author} {\bibfnamefont {W.}~\bibnamefont {Li}}, \bibinfo {author} {\bibfnamefont {F.}~\bibnamefont {Yang}}, \bibinfo {author} {\bibfnamefont {B.}~\bibnamefont {Chen}}, \ and\ \bibinfo {author} {\bibfnamefont {X.}~\bibnamefont {Xu}},\ }\bibfield  {title} {\enquote {\bibinfo {title} {Molecular dynamics simulation on the merging movement of nanodroplets on materials surface},}\ }\href@noop {} {\bibfield  {journal} {\bibinfo  {journal} {Results in Physics}\ }\textbf {\bibinfo {volume} {33}},\ \bibinfo {pages} {105213} (\bibinfo {year} {2022}{\natexlab{a}})}\BibitemShut {NoStop}%
\bibitem [{\citenamefont {Chang}, \citenamefont {Sheng},\ and\ \citenamefont {Tsao}(2016)}]{chang2016wetting}%
  \BibitemOpen
  \bibfield  {author} {\bibinfo {author} {\bibfnamefont {C.-C.}\ \bibnamefont {Chang}}, \bibinfo {author} {\bibfnamefont {Y.-J.}\ \bibnamefont {Sheng}}, \ and\ \bibinfo {author} {\bibfnamefont {H.-K.}\ \bibnamefont {Tsao}},\ }\bibfield  {title} {\enquote {\bibinfo {title} {Wetting hysteresis of nanodrops on nanorough surfaces},}\ }\href@noop {} {\bibfield  {journal} {\bibinfo  {journal} {Physical Review E}\ }\textbf {\bibinfo {volume} {94}},\ \bibinfo {pages} {042807} (\bibinfo {year} {2016})}\BibitemShut {NoStop}%
\bibitem [{\citenamefont {Chang}\ \emph {et~al.}(2016)\citenamefont {Chang}, \citenamefont {Wu}, \citenamefont {Sheng},\ and\ \citenamefont {Tsao}}]{chang2016resisting}%
  \BibitemOpen
  \bibfield  {author} {\bibinfo {author} {\bibfnamefont {C.-C.}\ \bibnamefont {Chang}}, \bibinfo {author} {\bibfnamefont {C.-J.}\ \bibnamefont {Wu}}, \bibinfo {author} {\bibfnamefont {Y.-J.}\ \bibnamefont {Sheng}}, \ and\ \bibinfo {author} {\bibfnamefont {H.-K.}\ \bibnamefont {Tsao}},\ }\bibfield  {title} {\enquote {\bibinfo {title} {Resisting and pinning of a nanodrop by trenches on a hysteresis-free surface},}\ }\href@noop {} {\bibfield  {journal} {\bibinfo  {journal} {The Journal of Chemical Physics}\ }\textbf {\bibinfo {volume} {145}} (\bibinfo {year} {2016})}\BibitemShut {NoStop}%
\bibitem [{\citenamefont {Chaudhri}\ \emph {et~al.}(2014)\citenamefont {Chaudhri}, \citenamefont {Bell}, \citenamefont {Garcia},\ and\ \citenamefont {Donev}}]{MultiphaseFHD2014}%
  \BibitemOpen
  \bibfield  {author} {\bibinfo {author} {\bibfnamefont {A.}~\bibnamefont {Chaudhri}}, \bibinfo {author} {\bibfnamefont {J.~B.}\ \bibnamefont {Bell}}, \bibinfo {author} {\bibfnamefont {A.~L.}\ \bibnamefont {Garcia}}, \ and\ \bibinfo {author} {\bibfnamefont {A.}~\bibnamefont {Donev}},\ }\bibfield  {title} {\enquote {\bibinfo {title} {Modeling multiphase flow using fluctuating hydrodynamics},}\ }\href {\doibase 10.1103/PhysRevE.90.033014} {\bibfield  {journal} {\bibinfo  {journal} {Phys. Rev. E}\ }\textbf {\bibinfo {volume} {90}},\ \bibinfo {pages} {033014} (\bibinfo {year} {2014})}\BibitemShut {NoStop}%
\bibitem [{\citenamefont {Klymko}\ \emph {et~al.}(2020)\citenamefont {Klymko}, \citenamefont {Nonaka}, \citenamefont {Bell}, \citenamefont {Carney},\ and\ \citenamefont {Garcia}}]{RTIL}%
  \BibitemOpen
  \bibfield  {author} {\bibinfo {author} {\bibfnamefont {K.}~\bibnamefont {Klymko}}, \bibinfo {author} {\bibfnamefont {A.}~\bibnamefont {Nonaka}}, \bibinfo {author} {\bibfnamefont {J.~B.}\ \bibnamefont {Bell}}, \bibinfo {author} {\bibfnamefont {S.~P.}\ \bibnamefont {Carney}}, \ and\ \bibinfo {author} {\bibfnamefont {A.~L.}\ \bibnamefont {Garcia}},\ }\bibfield  {title} {\enquote {\bibinfo {title} {Low mach number fluctuating hydrodynamics model for ionic liquids},}\ }\href {\doibase 10.1103/PhysRevFluids.5.093701} {\bibfield  {journal} {\bibinfo  {journal} {Phys. Rev. Fluids}\ }\textbf {\bibinfo {volume} {5}},\ \bibinfo {pages} {093701} (\bibinfo {year} {2020})}\BibitemShut {NoStop}%
\bibitem [{\citenamefont {Gallo}(2022)}]{Gallo2022}%
  \BibitemOpen
  \bibfield  {author} {\bibinfo {author} {\bibfnamefont {M.}~\bibnamefont {Gallo}},\ }\bibfield  {title} {\enquote {\bibinfo {title} {Thermal fluctuations in metastable fluids},}\ }\href {\doibase 10.1063/5.0132478} {\bibfield  {journal} {\bibinfo  {journal} {Physics of Fluids}\ }\textbf {\bibinfo {volume} {34}},\ \bibinfo {pages} {122011} (\bibinfo {year} {2022})}\BibitemShut {NoStop}%
\bibitem [{\citenamefont {Zhao}, \citenamefont {Sprittles},\ and\ \citenamefont {Lockerby}(2019)}]{RayPlatFHD}%
  \BibitemOpen
  \bibfield  {author} {\bibinfo {author} {\bibfnamefont {C.}~\bibnamefont {Zhao}}, \bibinfo {author} {\bibfnamefont {J.~E.}\ \bibnamefont {Sprittles}}, \ and\ \bibinfo {author} {\bibfnamefont {D.~A.}\ \bibnamefont {Lockerby}},\ }\bibfield  {title} {\enquote {\bibinfo {title} {Revisiting the {R}ayleigh–{P}lateau instability for the nanoscale},}\ }\href {\doibase 10.1017/jfm.2018.950} {\bibfield  {journal} {\bibinfo  {journal} {Journal of Fluid Mechanics}\ }\textbf {\bibinfo {volume} {861}},\ \bibinfo {pages} {R3} (\bibinfo {year} {2019})}\BibitemShut {NoStop}%
\bibitem [{\citenamefont {Landau}\ and\ \citenamefont {Lifshitz}(1959)}]{Landau_59}%
  \BibitemOpen
  \bibfield  {author} {\bibinfo {author} {\bibfnamefont {L.~D.}\ \bibnamefont {Landau}}\ and\ \bibinfo {author} {\bibfnamefont {E.~M.}\ \bibnamefont {Lifshitz}},\ }\href@noop {} {\emph {\bibinfo {title} {Fluid Mechanics, Course of Theoretical Physics, Vol. 6}}}\ (\bibinfo  {publisher} {Pergamon Press},\ \bibinfo {year} {1959})\BibitemShut {NoStop}%
\bibitem [{\citenamefont {de~Zarate}\ and\ \citenamefont {Sengers}(2007)}]{Zarate_07}%
  \BibitemOpen
  \bibfield  {author} {\bibinfo {author} {\bibfnamefont {J.~M.~O.}\ \bibnamefont {de~Zarate}}\ and\ \bibinfo {author} {\bibfnamefont {J.~V.}\ \bibnamefont {Sengers}},\ }\href@noop {} {\emph {\bibinfo {title} {Hydrodynamic Fluctuations in Fluids and Fluid Mixtures}}}\ (\bibinfo  {publisher} {Elsevier Science},\ \bibinfo {year} {2007})\BibitemShut {NoStop}%
\bibitem [{\citenamefont {Mecke}(2001)}]{ThinFilmMecke2001}%
  \BibitemOpen
  \bibfield  {author} {\bibinfo {author} {\bibfnamefont {K.~R.}\ \bibnamefont {Mecke}},\ }\bibfield  {title} {\enquote {\bibinfo {title} {Thermal fluctuations of thin liquid films},}\ }\href {\doibase 10.1088/0953-8984/13/21/302} {\bibfield  {journal} {\bibinfo  {journal} {Journal of Physics: Condensed Matter}\ }\textbf {\bibinfo {volume} {13}},\ \bibinfo {pages} {4615} (\bibinfo {year} {2001})}\BibitemShut {NoStop}%
\bibitem [{\citenamefont {Grün}, \citenamefont {Mecke},\ and\ \citenamefont {Rauscher}(2006)}]{ThinFilmGrun2006}%
  \BibitemOpen
  \bibfield  {author} {\bibinfo {author} {\bibfnamefont {G.}~\bibnamefont {Grün}}, \bibinfo {author} {\bibfnamefont {K.}~\bibnamefont {Mecke}}, \ and\ \bibinfo {author} {\bibfnamefont {M.}~\bibnamefont {Rauscher}},\ }\bibfield  {title} {\enquote {\bibinfo {title} {Thin-film flow influenced by thermal noise},}\ }\href@noop {} {\bibfield  {journal} {\bibinfo  {journal} {Journal of Statistical Physics}\ }\textbf {\bibinfo {volume} {122}},\ \bibinfo {pages} {1261} (\bibinfo {year} {2006})}\BibitemShut {NoStop}%
\bibitem [{\citenamefont {Zhang}, \citenamefont {Sprittles},\ and\ \citenamefont {Lockerby}(2021)}]{CapWaveThinFilm_2021}%
  \BibitemOpen
  \bibfield  {author} {\bibinfo {author} {\bibfnamefont {Y.}~\bibnamefont {Zhang}}, \bibinfo {author} {\bibfnamefont {J.}~\bibnamefont {Sprittles}}, \ and\ \bibinfo {author} {\bibfnamefont {D.}~\bibnamefont {Lockerby}},\ }\bibfield  {title} {\enquote {\bibinfo {title} {Thermal capillary wave growth and surface roughening of nanoscale liquid films},}\ }\href {\doibase 10.1017/jfm.2021.164} {\bibfield  {journal} {\bibinfo  {journal} {Journal of Fluid Mechanics}\ }\textbf {\bibinfo {volume} {915}},\ \bibinfo {pages} {A135} (\bibinfo {year} {2021})}\BibitemShut {NoStop}%
\bibitem [{\citenamefont {Zhao}\ \emph {et~al.}(2022{\natexlab{b}})\citenamefont {Zhao}, \citenamefont {Liu}, \citenamefont {Lockerby},\ and\ \citenamefont {Sprittles}}]{ThinFilmFluct_2021}%
  \BibitemOpen
  \bibfield  {author} {\bibinfo {author} {\bibfnamefont {C.}~\bibnamefont {Zhao}}, \bibinfo {author} {\bibfnamefont {J.}~\bibnamefont {Liu}}, \bibinfo {author} {\bibfnamefont {D.~A.}\ \bibnamefont {Lockerby}}, \ and\ \bibinfo {author} {\bibfnamefont {J.~E.}\ \bibnamefont {Sprittles}},\ }\bibfield  {title} {\enquote {\bibinfo {title} {Fluctuation-driven dynamics in nanoscale thin-film flows: Physical insights from numerical investigations},}\ }\href {\doibase 10.1103/PhysRevFluids.7.024203} {\bibfield  {journal} {\bibinfo  {journal} {Phys. Rev. Fluids}\ }\textbf {\bibinfo {volume} {7}},\ \bibinfo {pages} {024203} (\bibinfo {year} {2022}{\natexlab{b}})}\BibitemShut {NoStop}%
\bibitem [{\citenamefont {Liu}\ \emph {et~al.}(2023)\citenamefont {Liu}, \citenamefont {Zhao}, \citenamefont {Lockerby},\ and\ \citenamefont {Sprittles}}]{CapWaveFluct2023}%
  \BibitemOpen
  \bibfield  {author} {\bibinfo {author} {\bibfnamefont {J.}~\bibnamefont {Liu}}, \bibinfo {author} {\bibfnamefont {C.}~\bibnamefont {Zhao}}, \bibinfo {author} {\bibfnamefont {D.~A.}\ \bibnamefont {Lockerby}}, \ and\ \bibinfo {author} {\bibfnamefont {J.~E.}\ \bibnamefont {Sprittles}},\ }\bibfield  {title} {\enquote {\bibinfo {title} {Thermal capillary waves on bounded nanoscale thin films},}\ }\href {\doibase 10.1103/PhysRevE.107.015105} {\bibfield  {journal} {\bibinfo  {journal} {Phys. Rev. E}\ }\textbf {\bibinfo {volume} {107}},\ \bibinfo {pages} {015105} (\bibinfo {year} {2023})}\BibitemShut {NoStop}%
\bibitem [{\citenamefont {Cahn}\ and\ \citenamefont {Hilliard}(1958)}]{CahnHilliard:1958}%
  \BibitemOpen
  \bibfield  {author} {\bibinfo {author} {\bibfnamefont {J.~W.}\ \bibnamefont {Cahn}}\ and\ \bibinfo {author} {\bibfnamefont {J.}~\bibnamefont {Hilliard}},\ }\bibfield  {title} {\enquote {\bibinfo {title} {Free energy of a nonuniform system. i. interfacial free energy},}\ }\href@noop {} {\bibfield  {journal} {\bibinfo  {journal} {J. Chem. Phys}\ }\textbf {\bibinfo {volume} {28}} (\bibinfo {year} {1958})}\BibitemShut {NoStop}%
\bibitem [{\citenamefont {Anderson}, \citenamefont {McFadden},\ and\ \citenamefont {Wheeler}(1998)}]{anderson1998diffuse}%
  \BibitemOpen
  \bibfield  {author} {\bibinfo {author} {\bibfnamefont {D.~M.}\ \bibnamefont {Anderson}}, \bibinfo {author} {\bibfnamefont {G.~B.}\ \bibnamefont {McFadden}}, \ and\ \bibinfo {author} {\bibfnamefont {A.~A.}\ \bibnamefont {Wheeler}},\ }\bibfield  {title} {\enquote {\bibinfo {title} {Diffuse-interface methods in fluid mechanics},}\ }\href@noop {} {\bibfield  {journal} {\bibinfo  {journal} {Annual review of fluid mechanics}\ }\textbf {\bibinfo {volume} {30}},\ \bibinfo {pages} {139--165} (\bibinfo {year} {1998})}\BibitemShut {NoStop}%
\bibitem [{\citenamefont {Khatavkar}\ \emph {et~al.}(2007)\citenamefont {Khatavkar}, \citenamefont {Anderson}, \citenamefont {Duineveld},\ and\ \citenamefont {Meijer}}]{khatavkar2007diffuse}%
  \BibitemOpen
  \bibfield  {author} {\bibinfo {author} {\bibfnamefont {V.}~\bibnamefont {Khatavkar}}, \bibinfo {author} {\bibfnamefont {P.}~\bibnamefont {Anderson}}, \bibinfo {author} {\bibfnamefont {P.}~\bibnamefont {Duineveld}}, \ and\ \bibinfo {author} {\bibfnamefont {H.}~\bibnamefont {Meijer}},\ }\bibfield  {title} {\enquote {\bibinfo {title} {Diffuse-interface modelling of droplet impact},}\ }\href@noop {} {\bibfield  {journal} {\bibinfo  {journal} {Journal of Fluid Mechanics}\ }\textbf {\bibinfo {volume} {581}},\ \bibinfo {pages} {97--127} (\bibinfo {year} {2007})}\BibitemShut {NoStop}%
\bibitem [{\citenamefont {Heinen}\ \emph {et~al.}(2022)\citenamefont {Heinen}, \citenamefont {Hoffmann}, \citenamefont {Diewald}, \citenamefont {Seckler}, \citenamefont {Langenbach},\ and\ \citenamefont {Vrabec}}]{Droplets_MD_CH_2022}%
  \BibitemOpen
  \bibfield  {author} {\bibinfo {author} {\bibfnamefont {M.}~\bibnamefont {Heinen}}, \bibinfo {author} {\bibfnamefont {M.}~\bibnamefont {Hoffmann}}, \bibinfo {author} {\bibfnamefont {F.}~\bibnamefont {Diewald}}, \bibinfo {author} {\bibfnamefont {S.}~\bibnamefont {Seckler}}, \bibinfo {author} {\bibfnamefont {K.}~\bibnamefont {Langenbach}}, \ and\ \bibinfo {author} {\bibfnamefont {J.}~\bibnamefont {Vrabec}},\ }\bibfield  {title} {\enquote {\bibinfo {title} {{Droplet coalescence by molecular dynamics and phase-field modeling}},}\ }\href {\doibase 10.1063/5.0086131} {\bibfield  {journal} {\bibinfo  {journal} {Physics of Fluids}\ }\textbf {\bibinfo {volume} {34}},\ \bibinfo {pages} {042006} (\bibinfo {year} {2022})},\ \Eprint {http://arxiv.org/abs/https://pubs.aip.org/aip/pof/article-pdf/doi/10.1063/5.0086131/16600048/042006\_1\_online.pdf} {https://pubs.aip.org/aip/pof/article-pdf/doi/10.1063/5.0086131/16600048/042006\_1\_online.pdf} \BibitemShut {NoStop}%
\bibitem [{\citenamefont {Jacqmin}(2000)}]{ContactAngleBCs2000}%
  \BibitemOpen
  \bibfield  {author} {\bibinfo {author} {\bibfnamefont {D.}~\bibnamefont {Jacqmin}},\ }\bibfield  {title} {\enquote {\bibinfo {title} {Contact-line dynamics of a diffuse fluid interface},}\ }\href {\doibase 10.1017/S0022112099006874} {\bibfield  {journal} {\bibinfo  {journal} {Journal of Fluid Mechanics}\ }\textbf {\bibinfo {volume} {402}},\ \bibinfo {pages} {57–88} (\bibinfo {year} {2000})}\BibitemShut {NoStop}%
\bibitem [{\citenamefont {Dong}(2012)}]{ContactAngleBCs2012}%
  \BibitemOpen
  \bibfield  {author} {\bibinfo {author} {\bibfnamefont {S.}~\bibnamefont {Dong}},\ }\bibfield  {title} {\enquote {\bibinfo {title} {On imposing dynamic contact-angle boundary conditions for wall-bounded liquid–gas flows},}\ }\href {\doibase https://doi.org/10.1016/j.cma.2012.07.023} {\bibfield  {journal} {\bibinfo  {journal} {Computer Methods in Applied Mechanics and Engineering}\ }\textbf {\bibinfo {volume} {247-248}},\ \bibinfo {pages} {179--200} (\bibinfo {year} {2012})}\BibitemShut {NoStop}%
\bibitem [{\citenamefont {Seppecher}(1996)}]{Seppecher1996moving}%
  \BibitemOpen
  \bibfield  {author} {\bibinfo {author} {\bibfnamefont {P.}~\bibnamefont {Seppecher}},\ }\bibfield  {title} {\enquote {\bibinfo {title} {Moving contact lines in the cahn-hilliard theory},}\ }\href@noop {} {\bibfield  {journal} {\bibinfo  {journal} {International journal of engineering science}\ }\textbf {\bibinfo {volume} {34}},\ \bibinfo {pages} {977--992} (\bibinfo {year} {1996})}\BibitemShut {NoStop}%
\bibitem [{\citenamefont {Sibley}\ \emph {et~al.}(2013)\citenamefont {Sibley}, \citenamefont {Nold}, \citenamefont {Savva},\ and\ \citenamefont {Kalliadasis}}]{Sibley2013contactline}%
  \BibitemOpen
  \bibfield  {author} {\bibinfo {author} {\bibfnamefont {D.~N.}\ \bibnamefont {Sibley}}, \bibinfo {author} {\bibfnamefont {A.}~\bibnamefont {Nold}}, \bibinfo {author} {\bibfnamefont {N.}~\bibnamefont {Savva}}, \ and\ \bibinfo {author} {\bibfnamefont {S.}~\bibnamefont {Kalliadasis}},\ }\bibfield  {title} {\enquote {\bibinfo {title} {{The contact line behaviour of solid-liquid-gas diffuse-interface models}},}\ }\href {\doibase 10.1063/1.4821288} {\bibfield  {journal} {\bibinfo  {journal} {Physics of Fluids}\ }\textbf {\bibinfo {volume} {25}},\ \bibinfo {pages} {092111} (\bibinfo {year} {2013})}\BibitemShut {NoStop}%
\bibitem [{\citenamefont {Huh}\ and\ \citenamefont {Scriven}(1971)}]{Huh1971hydrodynamic}%
  \BibitemOpen
  \bibfield  {author} {\bibinfo {author} {\bibfnamefont {C.}~\bibnamefont {Huh}}\ and\ \bibinfo {author} {\bibfnamefont {L.~E.}\ \bibnamefont {Scriven}},\ }\bibfield  {title} {\enquote {\bibinfo {title} {Hydrodynamic model of steady movement of a solid/liquid/fluid contact line},}\ }\href@noop {} {\bibfield  {journal} {\bibinfo  {journal} {Journal of colloid and interface science}\ }\textbf {\bibinfo {volume} {35}},\ \bibinfo {pages} {85--101} (\bibinfo {year} {1971})}\BibitemShut {NoStop}%
\bibitem [{\citenamefont {Snoeijer}\ and\ \citenamefont {Andreotti}(2013)}]{Snoeijer2013moving}%
  \BibitemOpen
  \bibfield  {author} {\bibinfo {author} {\bibfnamefont {J.~H.}\ \bibnamefont {Snoeijer}}\ and\ \bibinfo {author} {\bibfnamefont {B.}~\bibnamefont {Andreotti}},\ }\bibfield  {title} {\enquote {\bibinfo {title} {Moving contact lines: scales, regimes, and dynamical transitions},}\ }\href@noop {} {\bibfield  {journal} {\bibinfo  {journal} {Annual review of fluid mechanics}\ }\textbf {\bibinfo {volume} {45}},\ \bibinfo {pages} {269--292} (\bibinfo {year} {2013})}\BibitemShut {NoStop}%
\bibitem [{\citenamefont {Barker}, \citenamefont {Bell},\ and\ \citenamefont {Garcia}(2023)}]{BrynRP2023}%
  \BibitemOpen
  \bibfield  {author} {\bibinfo {author} {\bibfnamefont {B.}~\bibnamefont {Barker}}, \bibinfo {author} {\bibfnamefont {J.~B.}\ \bibnamefont {Bell}}, \ and\ \bibinfo {author} {\bibfnamefont {A.~L.}\ \bibnamefont {Garcia}},\ }\bibfield  {title} {\enquote {\bibinfo {title} {Fluctuating hydrodynamics and the rayleigh–plateau instability},}\ }\href {\doibase 10.1073/pnas.2306088120} {\bibfield  {journal} {\bibinfo  {journal} {Proceedings of the National Academy of Sciences}\ }\textbf {\bibinfo {volume} {120}},\ \bibinfo {pages} {e2306088120} (\bibinfo {year} {2023})},\ \Eprint {http://arxiv.org/abs/https://www.pnas.org/doi/pdf/10.1073/pnas.2306088120} {https://www.pnas.org/doi/pdf/10.1073/pnas.2306088120} \BibitemShut {NoStop}%
\bibitem [{\citenamefont {Cai}\ \emph {et~al.}(2014)\citenamefont {Cai}, \citenamefont {Nonaka}, \citenamefont {Bell}, \citenamefont {Griffith},\ and\ \citenamefont {Donev}}]{cai:2014}%
  \BibitemOpen
  \bibfield  {author} {\bibinfo {author} {\bibfnamefont {M.}~\bibnamefont {Cai}}, \bibinfo {author} {\bibfnamefont {A.}~\bibnamefont {Nonaka}}, \bibinfo {author} {\bibfnamefont {J.~B.}\ \bibnamefont {Bell}}, \bibinfo {author} {\bibfnamefont {B.~E.}\ \bibnamefont {Griffith}}, \ and\ \bibinfo {author} {\bibfnamefont {A.}~\bibnamefont {Donev}},\ }\bibfield  {title} {\enquote {\bibinfo {title} {Efficient variable-coefficient finite-volume stokes solvers},}\ }\href@noop {} {\bibfield  {journal} {\bibinfo  {journal} {Communications in Computational Physics}\ }\textbf {\bibinfo {volume} {16}},\ \bibinfo {pages} {1263–1297} (\bibinfo {year} {2014})}\BibitemShut {NoStop}%
\bibitem [{\citenamefont {Donev}\ \emph {et~al.}(2014)\citenamefont {Donev}, \citenamefont {Nonaka}, \citenamefont {Sun}, \citenamefont {Fai}, \citenamefont {Garcia},\ and\ \citenamefont {Bell}}]{donev2014low}%
  \BibitemOpen
  \bibfield  {author} {\bibinfo {author} {\bibfnamefont {A.}~\bibnamefont {Donev}}, \bibinfo {author} {\bibfnamefont {A.}~\bibnamefont {Nonaka}}, \bibinfo {author} {\bibfnamefont {Y.}~\bibnamefont {Sun}}, \bibinfo {author} {\bibfnamefont {T.}~\bibnamefont {Fai}}, \bibinfo {author} {\bibfnamefont {A.}~\bibnamefont {Garcia}}, \ and\ \bibinfo {author} {\bibfnamefont {J.}~\bibnamefont {Bell}},\ }\bibfield  {title} {\enquote {\bibinfo {title} {Low mach number fluctuating hydrodynamics of diffusively mixing fluids},}\ }\href@noop {} {\bibfield  {journal} {\bibinfo  {journal} {Communications in Applied Mathematics and Computational Science}\ }\textbf {\bibinfo {volume} {9}},\ \bibinfo {pages} {47--105} (\bibinfo {year} {2014})}\BibitemShut {NoStop}%
\bibitem [{\citenamefont {Donev}\ \emph {et~al.}(2010)\citenamefont {Donev}, \citenamefont {Vanden-Eijnden}, \citenamefont {Garcia},\ and\ \citenamefont {Bell}}]{Donev_10}%
  \BibitemOpen
  \bibfield  {author} {\bibinfo {author} {\bibfnamefont {A.}~\bibnamefont {Donev}}, \bibinfo {author} {\bibfnamefont {E.}~\bibnamefont {Vanden-Eijnden}}, \bibinfo {author} {\bibfnamefont {A.~L.}\ \bibnamefont {Garcia}}, \ and\ \bibinfo {author} {\bibfnamefont {J.~B.}\ \bibnamefont {Bell}},\ }\bibfield  {title} {\enquote {\bibinfo {title} {On the accuracy of finite-volume schemes for fluctuating hydrodynamics},}\ }\href@noop {} {\bibfield  {journal} {\bibinfo  {journal} {Comm. Appl. Math and Comp. Sci.}\ }\textbf {\bibinfo {volume} {5}},\ \bibinfo {pages} {149} (\bibinfo {year} {2010})}\BibitemShut {NoStop}%
\bibitem [{\citenamefont {Bell}, \citenamefont {Nonaka},\ and\ \citenamefont {Garcia}(2024)}]{bell2024commentbrownianmotiondroplets}%
  \BibitemOpen
  \bibfield  {author} {\bibinfo {author} {\bibfnamefont {J.~B.}\ \bibnamefont {Bell}}, \bibinfo {author} {\bibfnamefont {A.}~\bibnamefont {Nonaka}}, \ and\ \bibinfo {author} {\bibfnamefont {A.~L.}\ \bibnamefont {Garcia}},\ }\href {https://arxiv.org/abs/2404.01444} {\enquote {\bibinfo {title} {Comment on "brownian motion of droplets induced by thermal noise"},}\ } (\bibinfo {year} {2024}),\ \Eprint {http://arxiv.org/abs/2404.01444} {arXiv:2404.01444 [cond-mat.soft]} \BibitemShut {NoStop}%
\bibitem [{\citenamefont {Happel}\ and\ \citenamefont {Brenner}(2012)}]{HappelBrenner2012}%
  \BibitemOpen
  \bibfield  {author} {\bibinfo {author} {\bibfnamefont {J.}~\bibnamefont {Happel}}\ and\ \bibinfo {author} {\bibfnamefont {H.}~\bibnamefont {Brenner}},\ }\href@noop {} {\emph {\bibinfo {title} {Low Reynolds number hydrodynamics: with special applications to particulate media}}},\ Vol.~\bibinfo {volume} {1}\ (\bibinfo  {publisher} {Springer Science \& Business Media},\ \bibinfo {year} {2012})\BibitemShut {NoStop}%
\bibitem [{\citenamefont {Hasimoto}(1959)}]{PeriodicLatticeDrag1959}%
  \BibitemOpen
  \bibfield  {author} {\bibinfo {author} {\bibfnamefont {H.}~\bibnamefont {Hasimoto}},\ }\bibfield  {title} {\enquote {\bibinfo {title} {On the periodic fundamental solutions of the stokes equations and their application to viscous flow past a cubic array of spheres},}\ }\href@noop {} {\bibfield  {journal} {\bibinfo  {journal} {Journal of Fluid Mechanics}\ }\textbf {\bibinfo {volume} {5}},\ \bibinfo {pages} {317--328} (\bibinfo {year} {1959})}\BibitemShut {NoStop}%
\bibitem [{\citenamefont {D{\"u}nweg}\ and\ \citenamefont {Kremer}(1993)}]{dunweg1993molecular}%
  \BibitemOpen
  \bibfield  {author} {\bibinfo {author} {\bibfnamefont {B.}~\bibnamefont {D{\"u}nweg}}\ and\ \bibinfo {author} {\bibfnamefont {K.}~\bibnamefont {Kremer}},\ }\bibfield  {title} {\enquote {\bibinfo {title} {Molecular dynamics simulation of a polymer chain in solution},}\ }\href@noop {} {\bibfield  {journal} {\bibinfo  {journal} {The Journal of chemical physics}\ }\textbf {\bibinfo {volume} {99}},\ \bibinfo {pages} {6983--6997} (\bibinfo {year} {1993})}\BibitemShut {NoStop}%
\bibitem [{\citenamefont {Li}(2009)}]{li2009BreakdownSE}%
  \BibitemOpen
  \bibfield  {author} {\bibinfo {author} {\bibfnamefont {Z.}~\bibnamefont {Li}},\ }\bibfield  {title} {\enquote {\bibinfo {title} {Critical particle size where the stokes-einstein relation breaks down},}\ }\href@noop {} {\bibfield  {journal} {\bibinfo  {journal} {Physical Review E—Statistical, Nonlinear, and Soft Matter Physics}\ }\textbf {\bibinfo {volume} {80}},\ \bibinfo {pages} {061204} (\bibinfo {year} {2009})}\BibitemShut {NoStop}%
\bibitem [{\citenamefont {Kawasaki}\ and\ \citenamefont {Kim}(2017)}]{Kawasaki2017BreakdownSE}%
  \BibitemOpen
  \bibfield  {author} {\bibinfo {author} {\bibfnamefont {T.}~\bibnamefont {Kawasaki}}\ and\ \bibinfo {author} {\bibfnamefont {K.}~\bibnamefont {Kim}},\ }\bibfield  {title} {\enquote {\bibinfo {title} {Identifying time scales for violation/preservation of stokes-einstein relation in supercooled water},}\ }\href {\doibase 10.1126/sciadv.1700399} {\bibfield  {journal} {\bibinfo  {journal} {Science Advances}\ }\textbf {\bibinfo {volume} {3}},\ \bibinfo {pages} {e1700399} (\bibinfo {year} {2017})}\BibitemShut {NoStop}%
\bibitem [{\citenamefont {Qin}\ \emph {et~al.}(2023)\citenamefont {Qin}, \citenamefont {Huang}, \citenamefont {Song},\ and\ \citenamefont {Xu}}]{DropletsNerstCH_2023}%
  \BibitemOpen
  \bibfield  {author} {\bibinfo {author} {\bibfnamefont {Y.}~\bibnamefont {Qin}}, \bibinfo {author} {\bibfnamefont {H.}~\bibnamefont {Huang}}, \bibinfo {author} {\bibfnamefont {Z.}~\bibnamefont {Song}}, \ and\ \bibinfo {author} {\bibfnamefont {S.}~\bibnamefont {Xu}},\ }\bibfield  {title} {\enquote {\bibinfo {title} {{Droplet dynamics: A phase-field model of mobile charges, polarization, and its leaky dielectric approximation}},}\ }\href {\doibase 10.1063/5.0159956} {\bibfield  {journal} {\bibinfo  {journal} {Physics of Fluids}\ }\textbf {\bibinfo {volume} {35}},\ \bibinfo {pages} {083327} (\bibinfo {year} {2023})},\ \Eprint {http://arxiv.org/abs/https://pubs.aip.org/aip/pof/article-pdf/doi/10.1063/5.0159956/18094272/083327\_1\_5.0159956.pdf} {https://pubs.aip.org/aip/pof/article-pdf/doi/10.1063/5.0159956/18094272/083327\_1\_5.0159956.pdf} \BibitemShut {NoStop}%
\bibitem [{\citenamefont {P\'eraud}\ \emph {et~al.}(2016)\citenamefont {P\'eraud}, \citenamefont {Nonaka}, \citenamefont {Chaudhri}, \citenamefont {Bell}, \citenamefont {Donev},\ and\ \citenamefont {Garcia}}]{peraud2016}%
  \BibitemOpen
  \bibfield  {author} {\bibinfo {author} {\bibfnamefont {J.-P.}\ \bibnamefont {P\'eraud}}, \bibinfo {author} {\bibfnamefont {A.}~\bibnamefont {Nonaka}}, \bibinfo {author} {\bibfnamefont {A.}~\bibnamefont {Chaudhri}}, \bibinfo {author} {\bibfnamefont {J.~B.}\ \bibnamefont {Bell}}, \bibinfo {author} {\bibfnamefont {A.}~\bibnamefont {Donev}}, \ and\ \bibinfo {author} {\bibfnamefont {A.~L.}\ \bibnamefont {Garcia}},\ }\bibfield  {title} {\enquote {\bibinfo {title} {Low mach number fluctuating hydrodynamics for electrolytes},}\ }\href@noop {} {\bibfield  {journal} {\bibinfo  {journal} {Phys. Rev. Fluids}\ }\textbf {\bibinfo {volume} {1}},\ \bibinfo {pages} {074103} (\bibinfo {year} {2016})}\BibitemShut {NoStop}%
\bibitem [{\citenamefont {{Erik Teigen}}\ \emph {et~al.}(2011)\citenamefont {{Erik Teigen}}, \citenamefont {Song}, \citenamefont {Lowengrub},\ and\ \citenamefont {Voigt}}]{Teigen2011surfactant}%
  \BibitemOpen
  \bibfield  {author} {\bibinfo {author} {\bibfnamefont {K.}~\bibnamefont {{Erik Teigen}}}, \bibinfo {author} {\bibfnamefont {P.}~\bibnamefont {Song}}, \bibinfo {author} {\bibfnamefont {J.}~\bibnamefont {Lowengrub}}, \ and\ \bibinfo {author} {\bibfnamefont {A.}~\bibnamefont {Voigt}},\ }\bibfield  {title} {\enquote {\bibinfo {title} {A diffuse-interface method for two-phase flows with soluble surfactants},}\ }\href {\doibase https://doi.org/10.1016/j.jcp.2010.09.020} {\bibfield  {journal} {\bibinfo  {journal} {Journal of Computational Physics}\ }\textbf {\bibinfo {volume} {230}},\ \bibinfo {pages} {375--393} (\bibinfo {year} {2011})}\BibitemShut {NoStop}%
\bibitem [{\citenamefont {Zong}\ \emph {et~al.}(2020)\citenamefont {Zong}, \citenamefont {Zhang}, \citenamefont {Liang}, \citenamefont {Wang},\ and\ \citenamefont {Xu}}]{Zong2020surfactant}%
  \BibitemOpen
  \bibfield  {author} {\bibinfo {author} {\bibfnamefont {Y.}~\bibnamefont {Zong}}, \bibinfo {author} {\bibfnamefont {C.}~\bibnamefont {Zhang}}, \bibinfo {author} {\bibfnamefont {H.}~\bibnamefont {Liang}}, \bibinfo {author} {\bibfnamefont {L.}~\bibnamefont {Wang}}, \ and\ \bibinfo {author} {\bibfnamefont {J.}~\bibnamefont {Xu}},\ }\bibfield  {title} {\enquote {\bibinfo {title} {Modeling surfactant-laden droplet dynamics by lattice boltzmann method},}\ }\href@noop {} {\bibfield  {journal} {\bibinfo  {journal} {Physics of Fluids}\ }\textbf {\bibinfo {volume} {32}} (\bibinfo {year} {2020})}\BibitemShut {NoStop}%
\bibitem [{\citenamefont {Arbabi}\ \emph {et~al.}(2023)\citenamefont {Arbabi}, \citenamefont {Deuar}, \citenamefont {Denys}, \citenamefont {Bennacer}, \citenamefont {Che},\ and\ \citenamefont {Theodorakis}}]{Arbabi2023surfactant}%
  \BibitemOpen
  \bibfield  {author} {\bibinfo {author} {\bibfnamefont {S.}~\bibnamefont {Arbabi}}, \bibinfo {author} {\bibfnamefont {P.}~\bibnamefont {Deuar}}, \bibinfo {author} {\bibfnamefont {M.}~\bibnamefont {Denys}}, \bibinfo {author} {\bibfnamefont {R.}~\bibnamefont {Bennacer}}, \bibinfo {author} {\bibfnamefont {Z.}~\bibnamefont {Che}}, \ and\ \bibinfo {author} {\bibfnamefont {P.~E.}\ \bibnamefont {Theodorakis}},\ }\bibfield  {title} {\enquote {\bibinfo {title} {{Coalescence of surfactant-laden droplets}},}\ }\href {\doibase 10.1063/5.0153676} {\bibfield  {journal} {\bibinfo  {journal} {Physics of Fluids}\ }\textbf {\bibinfo {volume} {35}},\ \bibinfo {pages} {063329} (\bibinfo {year} {2023})}\BibitemShut {NoStop}%
\end{thebibliography}%

\end{document}